\documentclass[aps,pra,superscriptaddress,floatfix,longbibliography,onecolumn]{revtex4-2}
\usepackage{dcolumn}
\usepackage{graphicx,color}
\usepackage{amsmath,amssymb,bm,amsfonts,dsfont,mathrsfs,amsthm}
\usepackage{braket}
\usepackage{comment}
\usepackage{booktabs}
\usepackage{multirow}
\usepackage{siunitx}
\usepackage[normalem]{ulem}

\usepackage{enumitem}
\usepackage[colorlinks=true, allcolors=blue!62!black]{hyperref}

\usepackage{braket}          

\usepackage{subcaption}
\usepackage{makecell}

\usepackage{orcidlink}       
\usepackage[most]{tcolorbox}

\newtcolorbox{wideexplanationbox}[1]{
    enhanced,
    breakable,
    colback=black!2,
    colframe=black!65,
    colbacktitle=black!8,
    coltitle=black,
    fonttitle=\bfseries,
    title={#1},
    boxrule=0.5pt,
    arc=1mm,
    left=10pt,
    right=10pt,
    top=8pt,
    bottom=8pt,
    before skip=10pt,
    after skip=10pt
}

\usepackage{lipsum}          

\usepackage{ragged2e}
\usepackage[percent]{overpic}
\begin{document}

	\title{Nonlinear Function Regression Using Quantum Ensemble Learning }
	
	\author{Yueen Chen\,\orcidlink{0009-0003-6389-8954}
	}
	\email[]{chenyueen@uestc.edu.cn}
	\affiliation{Institute of Fundamental and Frontier Sciences, University of Electronic Science and Technology of China, Chengdu 611731, China}
	\affiliation{Key Laboratory of Quantum Physics and Photonic Quantum Information, Ministry of Education, University of Electronic Science and Technology of China, Chengdu 611731 , China}
	
	\author{Mohammad Sharifian\,\orcidlink{0000-0001-9801-239x}
	}
	\email[]{mohammad.sharifian@uestc.edu.cn}
	\affiliation{Institute of Fundamental and Frontier Sciences, University of Electronic Science and Technology of China, Chengdu 611731, China}
	\affiliation{Key Laboratory of Quantum Physics and Photonic Quantum Information, Ministry of Education, University of Electronic Science and Technology of China, Chengdu 611731 , China}
	
	
	\author{Abolfazl Bayat\,\orcidlink{0000-0003-3852-4558}
	}
	\email[]{abolfazl.bayat@uestc.edu.cn}
	\affiliation{Institute of Fundamental and Frontier Sciences, University of Electronic Science and Technology of China, Chengdu 611731, China}
	\affiliation{Key Laboratory of Quantum Physics and Photonic Quantum Information, Ministry of Education, University of Electronic Science and Technology of China, Chengdu 611731 , China}
	\affiliation{Shimmer Center, Tianfu Jiangxi Laboratory, Chengdu 641419, China}
	
\begin{abstract}
Recent advances in artificial intelligence  and quantum technologies have driven the emergence of quantum machine learning, yet imperfections in near-term quantum devices remain a major obstacle for achieving practical quantum advantage. In the absence of fault-tolerant machines, ensemble learning, which combines multiple imperfect quantum devices, may unlock new quantum computing applications. Here, we introduce two ensemble-learning algorithms, namely Bagging and AdaBoost, for enhancing the accuracy of nonlinear regression problems. While Bagging trains variational quantum regressors independently, AdaBoost employs a sequential training strategy and achieves significantly better prediction accuracy. By incorporating systematic feature selection into our ensemble learning algorithms, we enhance their predictive power for highly nonlinear functions. We further improve prediction accuracy through a feature-to-qubit mapping strategy based on repeated feature encoding. The performance of our ensemble-based algorithms improves with both the number of regressors in the ensemble and the size of each regressor. We demonstrate the effectiveness of our approach using four benchmarking functions with varying degrees of nonlinearity by relying on a feature pool limited to first- and second-order polynomial terms.

\end{abstract}
	
\maketitle

\section{Introduction}

The superiority of quantum computation over its classical counterpart has been proven through several  algorithms for solving certain problems, such as integer factoring~\cite{shor1994algorithms}, search problems~\cite{grover1996fast}, and solving linear equations~\cite{harrow2009quantum} which inspired several other algorithms~\cite{wiebe2012quantum,Wang2014quantum,schuld2016prediction,Liu2017fast}. However, implementation of those quantum algorithms require fault-tolerant quantum computers which are not yet available. The current Noisy Intermediate-Scale Quantum (NISQ) devices suffer from various imperfections in their initialization, operation and readout~\cite{endo2021hybrid,
temme2017error,kandala2019error,krinner2019engineering,
kjaergaard2020superconducting,saffman2016quantum,kaufman2021quantum}. Heuristic quantum algorithms, whose quantum advantage are yet to be proven, have been developed for implementation on NISQ simulators, providing proof of principle results and exploring potential applications for quantum computation~\cite{preskill2018quantum,bharti2022noisy,zimboras2025myths,cerezo2021variational,fujii2017harnessing,Xia_2023,hu2019quantum,nakanishi2019subspace,cao2026theory,sannia2026temporal,Sharifian2026quantum,das2026quatum}. Variational quantum algorithms constitute an important class of heuristic algorithms that hybridize quantum computation with classical optimization, thereby distributing the complexity between classical and quantum computers~\cite{peruzzo2014variational,
cerezo2021variational,mcclean2016theory,benedetti2019parameterized,Sharifian2025hybrid,sweke2020stochastic}. The architecture and optimization of parameterized quantum circuits can
also be assisted by classical learning models and architecture-search
strategies~\cite{zhang2021neural,miao2024neural}. Such algorithms have been used for a wide range of applications in quantum chemistry~\cite{kandala2017hardware,
cao2019quantum,mcardle2020quantum,tilly2022variational}, machine learning~\cite{biamonte2017quantum,dunjko2018machine,
2018PhRvA..98c2309M,schuld2019quantum,havlivcek2019supervised,
huang2021power,liu2021rigorous,jerbi2023quantum,
cerezo2022challenges}, optimization~\cite{farhi2014quantum,zhou2020quantum,hadfield2019quantum,harrigan2021quantum} and condensed matter physics~\cite{li2017efficient,yuan2019theory,
kokail2019self,carleo2017solving}. Within quantum machine learning, circuit architecture
strongly affects model expressibility~\cite{sim2019expressibility}. Generalization
and robustness to adversarial perturbations and quantum
noise have also become
important research questions~\cite{banchi2021generalization,gil2024understanding,liu2020vulnerability,du2021quantum}. Nonetheless, variational quantum algorithms face sever challenges in their training due to infamous barren plateau~\cite{mcclean2018barren,
2021NatCo..12.1791C,wang2021noise,holmes2022connecting,ortiz2021entanglement}, making them limited to shallow circuits. An important question is whether one can compensate for the weak performance of shallow circuits in such quantum algorithms by exploiting an ensemble of them.\\

Indeed, quantum ensemble learning algorithms have emerged as NISQ-friendly protocols to address the above problem. Two prominent ensemble-based quantum algorithms are: (i) bootstrap aggregating (which is commonly known as Bagging); and (ii) adaptive boosting (which is commonly known as AdaBoost).    
Bagging algorithm combines independently trained models to improve predictive stability~\cite{breiman1996bagging,
buhlmann2002analyzing}. In contrast, AdaBoost sequentially trains weak predictors to gradually enhance their performance~\cite{schapire1990strength,
freund1997decision,drucker1997improvingRU}. In each sequence, those data samples which are wrongly predicted systematically gain more weight during the training of the next predictor forcing it to predict them more precisely. This sequential training may improve the quality of prediction at each step. Both Bagging and AdaBoost algorithms have been used for addressing classification problems in the context of variational quantum algorithms~\cite{schuld2018quantum,2023arXiv230311283I,
Li2024ensemble,wang2024supervised,chen2025quantum,Li2024ensemble}. \\

Regression is a fundamental machine learning task that aims to predict numerical outputs from input data by learning the relationship between input variables~\cite{hastie2009elements,bishop2006pattern}. The output of the predictor depends on the input data as well as a few trainable parameters of the regression function. In linear regression, the dependence of the output on the trainable parameters is linear and the input data are directly used in the mapping. If the relationship between input variables and target values is highly nonlinear then linear regression fails to provide accurate prediction. In this situation, it is better to use a regression function which depends nonlinearly on the trainable parameters. The function can also depend nonlinearly on  the input data to be more versatile in its prediction. However, in this situation, there are literally many possibilities for making new features via nonlinear combination of input data, e.g. high order polynomials~\cite{hastie2009elements} or Fourier basis~\cite{rahimi2007random,2021PhRvA.103c2430S}. Variational quantum algorithms have been used for making quantum regressors~\cite{wiebe2012quantum,schuld2016prediction,wang2017quantum,perez2021one,watabe2021quantum,hatakeyama2023quantum,kobayashi2023cross,panadero2024regressions,2023arXiv230703334W,souza2024regression,2023arXiv230311283I}, however, there are still several issues to be addressed. First, a systematic analysis for feature selection and its impact on the regression performance is missing~\cite{ho1998random,breiman2001random,schuld2019quantum,
2021PhRvA.103c2430S,perez2020data}. Second, nonlinear function prediction via quantum regression has not yet been systematically investigated across functions with different degrees of nonlinearity. Third, the prediction often has been made based on measuring a subset of qubits~\cite{2018PhRvA..98c2309M,watabe2021quantum,
hatakeyama2023quantum,panadero2024regressions}, sometimes even a single qubit~\cite{perez2021one,souza2024regression}, and thus a lot of information has been wasted at the output of the circuit.\\

In this work, we address regression problem through systematic incorporating of feature selection into quantum ensemble learning using variational quantum circuits. Specifically, we focus on two ensemble-based algorithms namely Bagging and AdaBoost.   We consider four benchmark functions with different degrees of nonlinearity, each depending on four input variables. A pool of 14 first- and second-order polynomial features, comprising linear, squared, and pairwise interaction terms are used as the input feature set. Our findings can be summarized as follows. First, we show that using the measurement outcomes of all qubits can improve regression accuracy by extracting more information from the quantum circuit. Second, while Bagging improves performance compared with individual regressors, the proposed greedy AdaBoost algorithm, which combines adaptive feature selection with sample-weight updates, achieves a greater improvement in regression accuracy. Third, we show that repeating selected features on two neighboring qubits within the greedy AdaBoost framework enhances the nonlinear response of the circuit and improves prediction accuracy without explicitly introducing higher-order features into the feature pool. Fourth,  increasing the number of qubits and the number of regressors  in the ensemble  monotonically improves the predictive accuracy showing that scaling up in both physical and ensemble sizes lead to  better performance.

\section{Data Generation and Quantum Regressor}
	\label{sec:regressor}
	
	Variational quantum regressor provide a parameterized quantum circuit approach for learning various mappings. One of the essential tools for modeling complex nonlinear systems and identifying unknown input-output relations from observed data is function approximation. In this work, we employ regressors to approximate a set of target functions exhibiting different degrees and forms of nonlinearity. This section introduces the regression tasks and their corresponding datasets, followed by the variational quantum circuit that serves as the fundamental building block of the ensemble models suggested in the subsequent sections.

	\subsection{Target Functions}
	
\begin{figure*}[t]
		\centering
		\captionsetup[subfigure]{skip=0pt}
		
		\begin{subfigure}[t]{0.245\textwidth}
			\begin{overpic}[width=\textwidth]{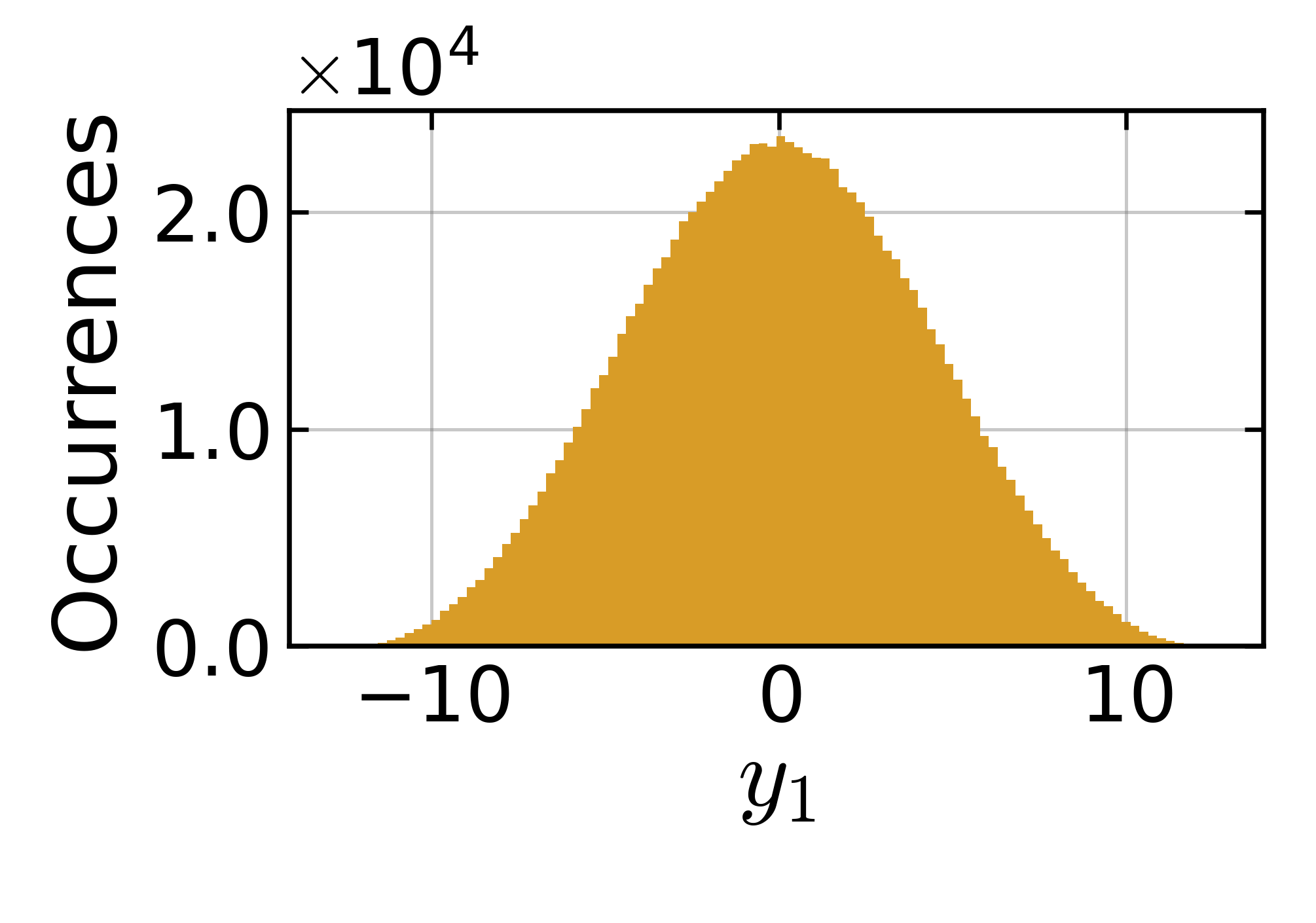}
				\put(3,65){(a)}
			\end{overpic}
			\label{fig:mult2}
		\end{subfigure}%
		\hspace{0.003\textwidth}%
		\begin{subfigure}[t]{0.245\textwidth}
			\begin{overpic}[width=\textwidth]{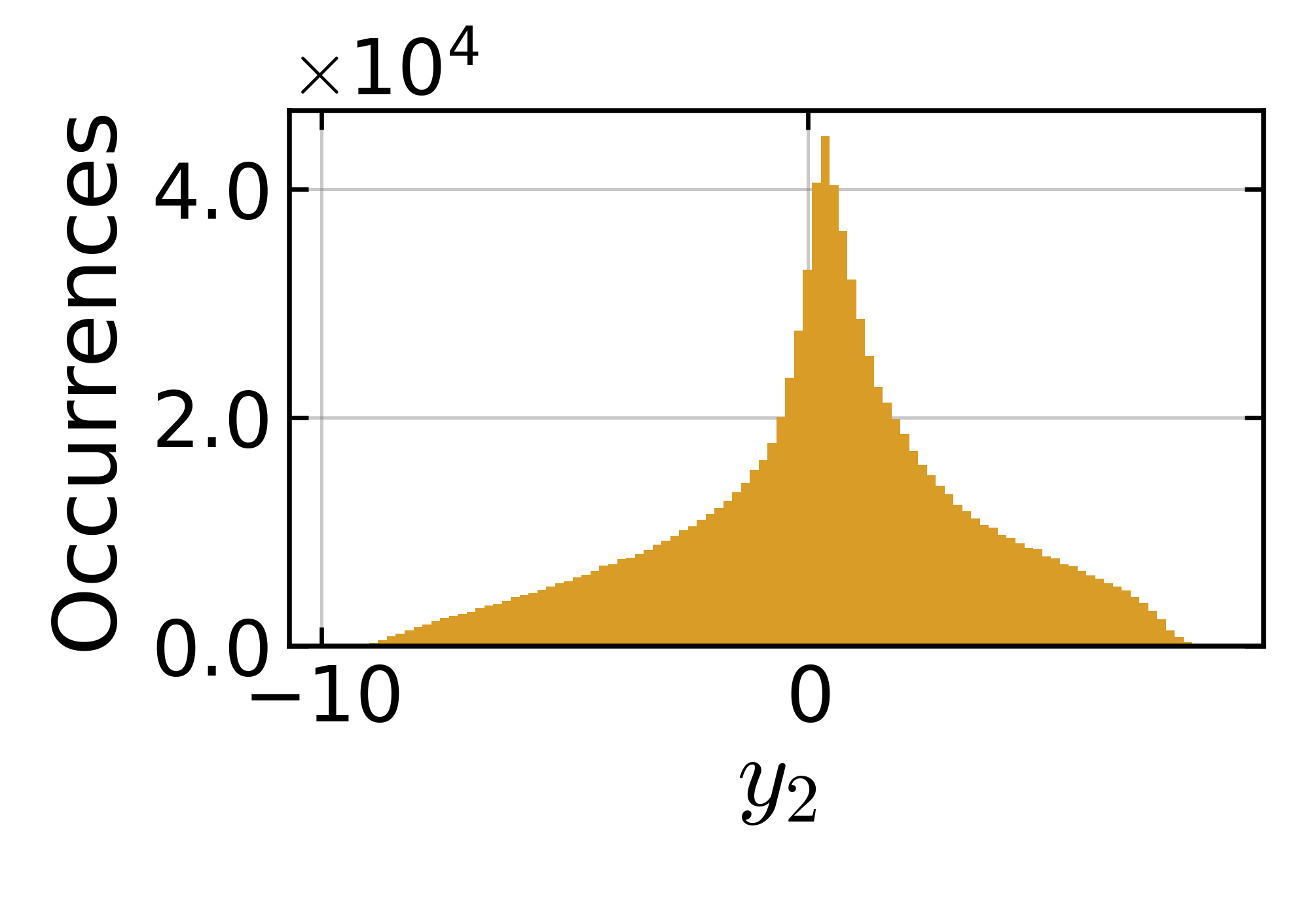}
				\put(3,65){(b)}
			\end{overpic}
			\label{fig:mult4}
		\end{subfigure}%
		\hspace{0.003\textwidth}%
		\begin{subfigure}[t]{0.245\textwidth}
			\begin{overpic}[width=\textwidth]{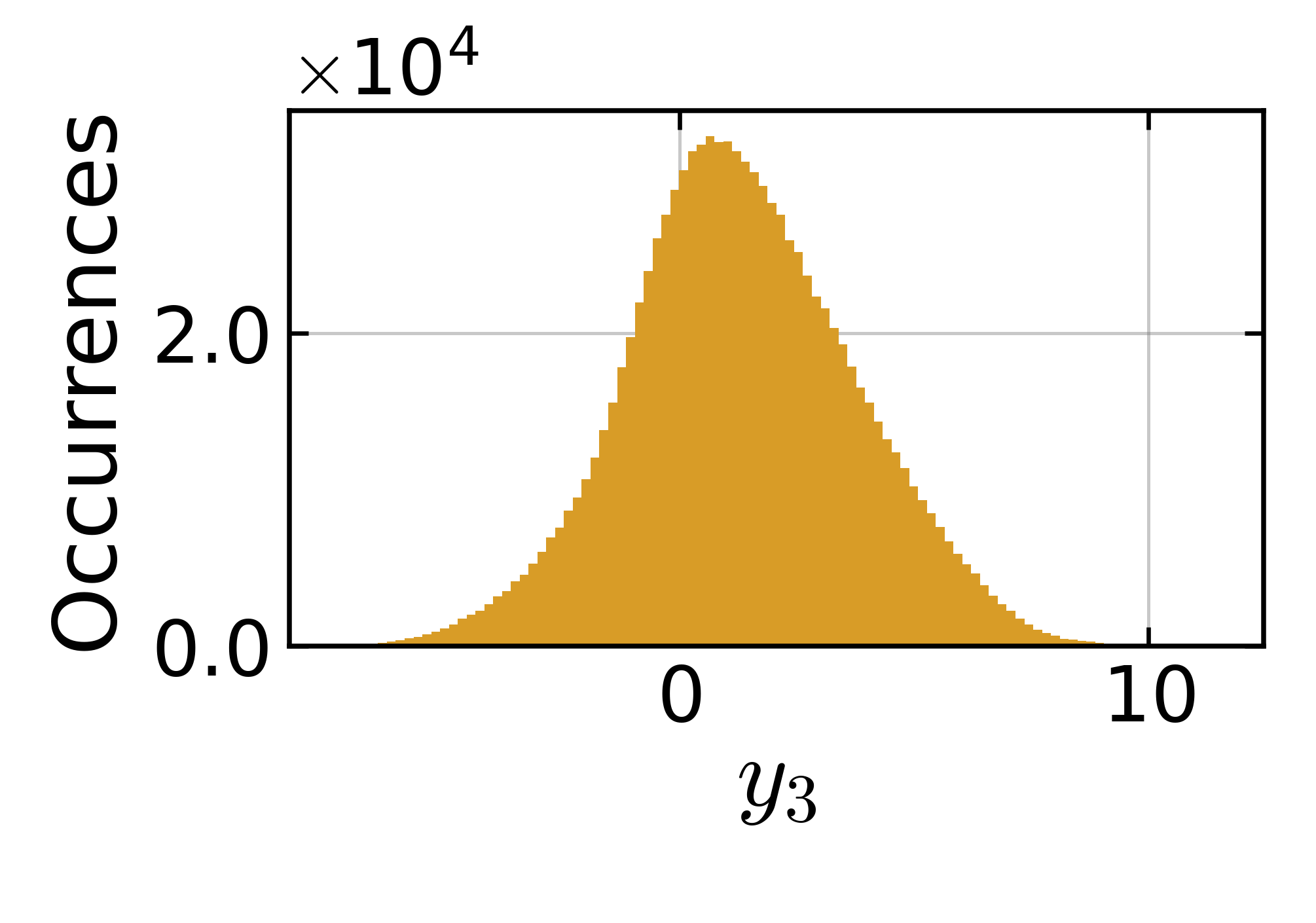}
				\put(3,65){(c)}
			\end{overpic}
			\label{fig:mult5}
		\end{subfigure}%
		\hspace{0.003\textwidth}%
		\begin{subfigure}[t]{0.245\textwidth}
			\begin{overpic}[width=\textwidth]{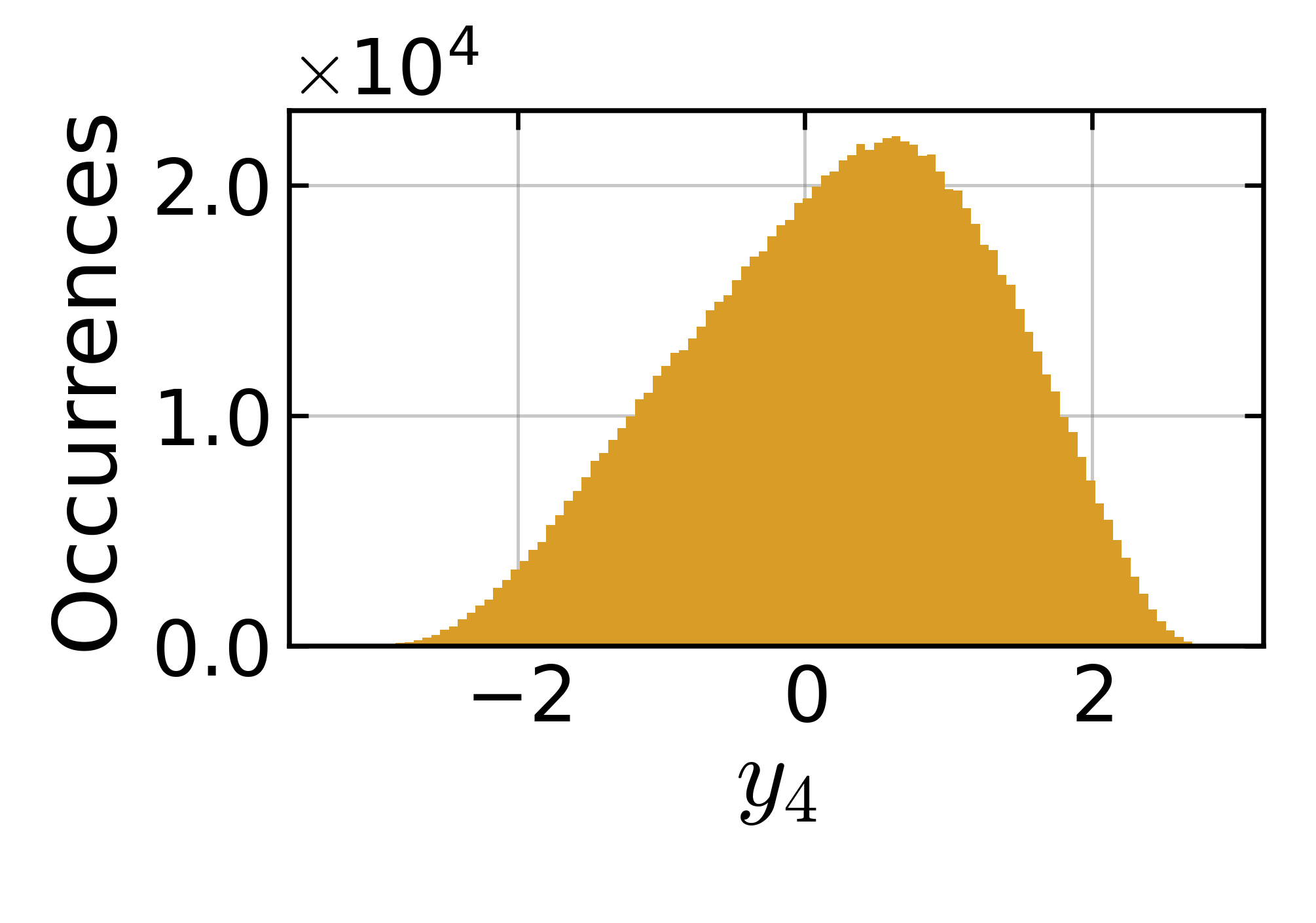}
				\put(3,65){(d)}
			\end{overpic}
			\label{fig:mult6}
		\end{subfigure}
		\vspace{-20pt}
		\caption{Output distributions of the four target mappings evaluated on one million input samples. Panels (a)--(d) correspond to $y_1$, $y_2$, $y_3$, and $y_4$, respectively. }
		\label{fig:target_distributions}
	\end{figure*}

	We consider a four dimensional input vector
	$\mathbf{x}{=}[x_1,x_2,x_3,x_4]^{\mathsf T}$,
	whose components are independently sampled from the uniform
	distribution $x_j{\sim}\mathcal{U}(-1,1)$, where $j{=}1,{\ldots},4$.
	Using these inputs, we define four target functions with distinct
	mathematical structures, including linear, quadratic, cubic
	polynomial, and fractional-power nonlinear mappings, respectively, as
	\begin{align}
		y_1(\mathbf{x})={}&x_1-3x_2+4x_3-5x_4,
		\label{eq:y1} \\
		y_2(\mathbf{x})={}&3x_1^2-2x_2^2+4x_1x_2-x_3x_4+4x_1-2x_2,
		\label{eq:y2} \\
		y_3(\mathbf{x})={}&3x_1^3-2x_1x_2^2+x_3^2x_4+x_1x_4^2
		+4x_2^2+x_1x_3-3x_2x_3+x_1-2x_4,
		\label{eq:y3}\\
		y_4(\mathbf{x})={}&2|x_1|^{2/3}-3|x_2|^{3/2}
		+|x_3|^{1/4}-|x_4|^{1/2}
		+2|x_2|^{3/2}|x_4|^2
		+|x_3|^{1/3}|x_4|
		-|x_1|^{1/4}|x_2|^{1/2}.
		\label{eq:y4}
	\end{align}
	For the $i$-th input sample, we write $\mathbf{x}_i{=}[x_{i,1},x_{i,2},x_{i,3},x_{i,4}]^{\mathsf T}$, where $i{=}1,\ldots,N$ denotes the sample index and $x_{i,j}$ denotes its $j$-th component. 
    For each input sample $\mathbf{x}_i$, the corresponding target value for the $k$-th mapping is $y_k(\mathbf{x}_i)$, where $i$ denotes the sample index, $i{=}1,\ldots,N$, and $k{=}1,\ldots,4$. The above four target functions serve as benchmark regression tasks to evaluate the performance of our quantum regressors across different mathematical structures and to allow us to investigate the effects of feature selection and ensemble learning on predictive performance. The first function, \(y_1\), evaluates the ability of the quantum regressor to approximate linear mappings. The second and third functions, \(y_2\) and \(y_3\), represent quadratic and cubic polynomial mappings, respectively. The fourth function, \(y_4\), contains fractional-power terms and exhibits more complex, nonpolynomial nonlinear behavior, providing a challenging benchmark for evaluating the model's ability to approximate nonlinear input-output relationships. Fig.~\ref{fig:target_distributions} shows the probability distributions of the outputs $y_1,y_2,y_3$, and $y_4$ obtained from one million independently sampled input vectors $\mathbf{x}_i$. Although all four regression tasks are constructed from identically distributed input variables, their distinct functional forms give rise to different output distributions. For instance, while the linear function $y_1$ obeys a Gaussian distribution the other functions show highly non-Gaussian distributions.   
    Together with their distinct mathematical structures, these different distributions indicate that the four functions present diverse regression tasks, making them suitable benchmarks for evaluating the predictive performance of the proposed models.\\
	
	For model training and evaluation, each regressor uses $N{=}1000$ samples, partitioned into training, validation, and test sets in the proportions $64\%$, $16\%$, and $20\%$, respectively.  While the data is four-dimensional, we define a set of feature  pool that combines the input parameters $x_j$'s. The elements of this feature set serve as inputs to our quantum regressors. Specifically, we construct a feature set comprising all first- and second-order polynomial terms of the four input variables as:
	\begin{equation}
		\Phi(\mathbf{x})=
		\bigl\{
		x_1,x_2,x_3,x_4,
		x_1^2,x_2^2,x_3^2,x_4^2,
		x_1x_2,x_1x_3,x_1x_4,
		x_2x_3,x_2x_4,x_3x_4
		\bigr\}.
		\label{eq:feature_pool}
	\end{equation}
	The resulting \(|\Phi(\mathbf{x})|{=}14\) features comprise four first-order terms, four squared terms, and six pairwise interaction terms. This set serves as a feature pool for the regression approaches presented in the following sections, where the input features for each regressor are selected from it.

	\subsection{Quantum regressor model}
	
\begin{figure*}[!t]
		\centering
		
		\begin{subfigure}[c]{0.7\textwidth}
			\centering
			\includegraphics[
			width=\linewidth,
			keepaspectratio
			]{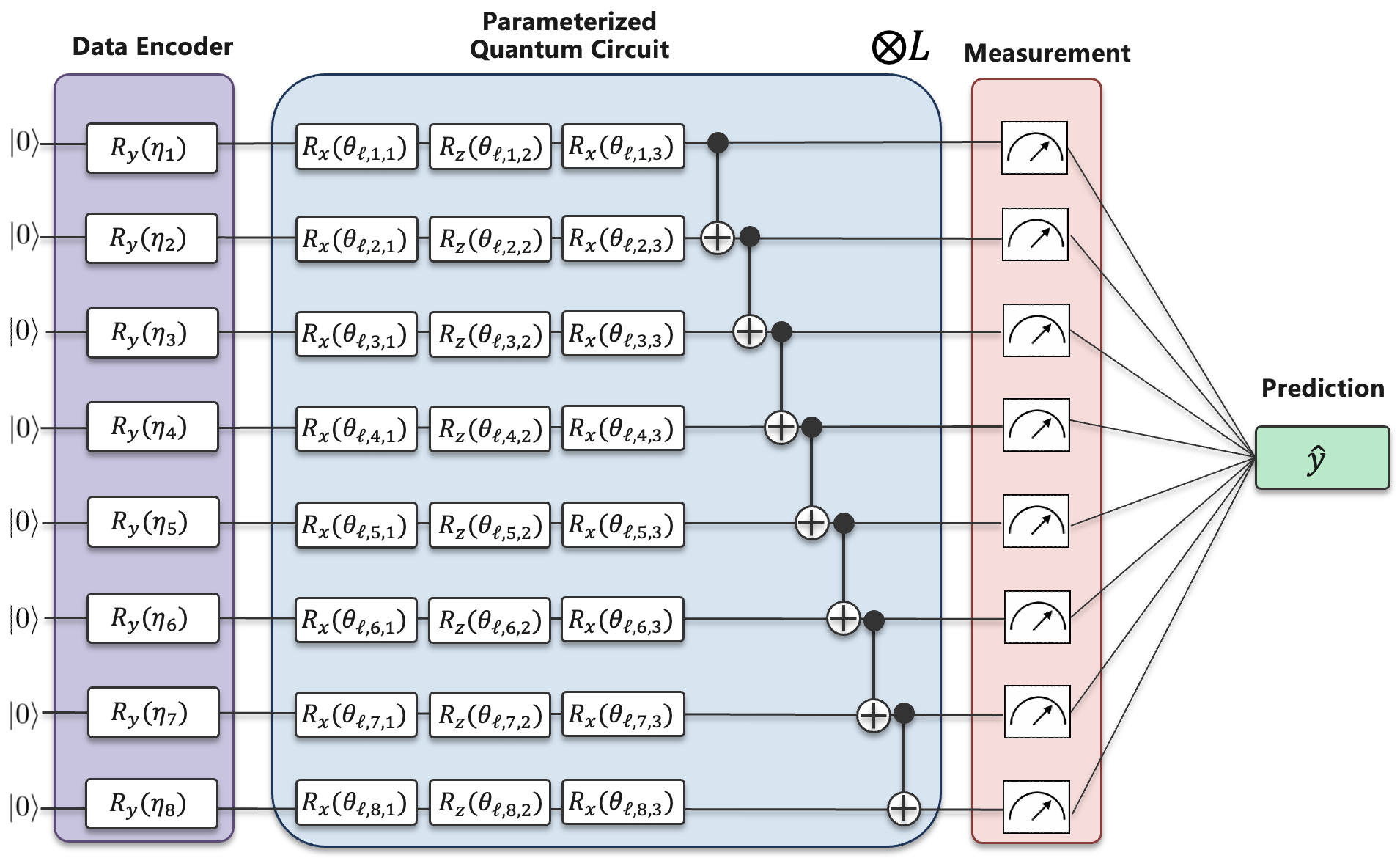}
			\put(-370,230){{(a)}}
			\label{fig:algo_bag1}
		\end{subfigure}
		\hfill
		\begin{subfigure}[c]{0.28\textwidth}
			\centering
			\includegraphics[
			width=\linewidth,
			height=0.36\textheight,
			keepaspectratio
			]{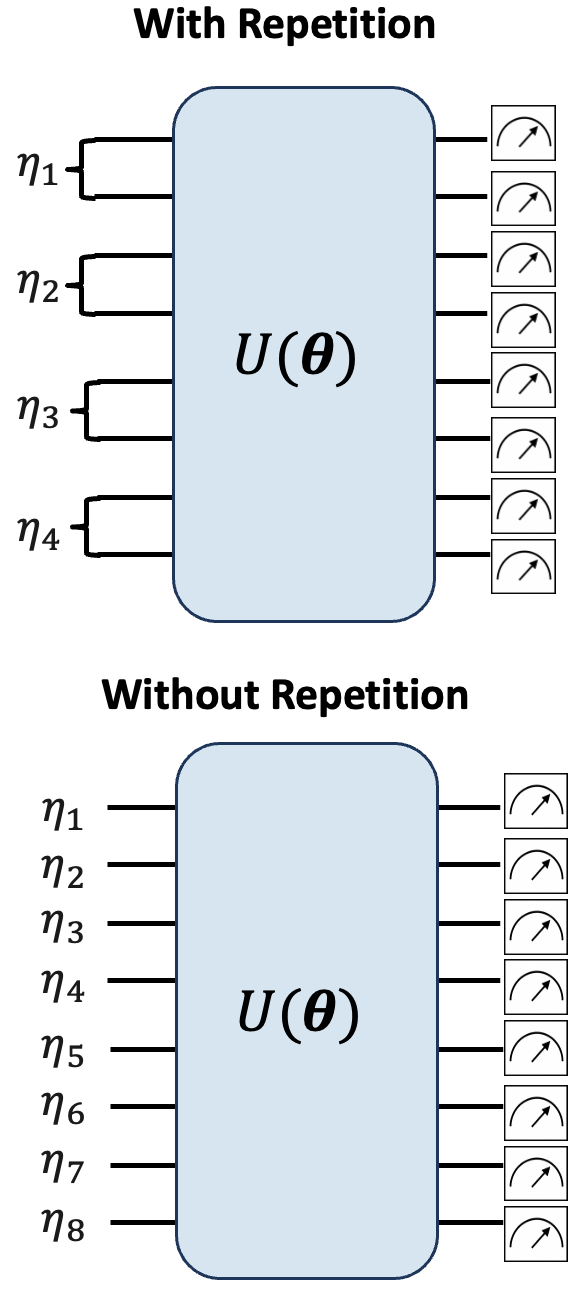}
			\put(-150,230){{(b)}}
			\label{fig:algo_bag2}
		\end{subfigure}
		
		\caption{(a) Schematic of the quantum regressor used in this work, consisting of three parts: feature encoding, parameterized quantum circuit, and classical readout. The parameterized circuit can be repeated $L$ times to increase the circuit depth. (b) Two feature-encoding strategies considered in the main text. With repetition, each selected feature is encoded onto a pair of adjacent qubits, whereas without repetition, each qubit encodes a distinct feature.}
		\label{fig:bagging_workflow}
	\end{figure*}

	In this section, we explain how we are going to make quantum regressors using variational quantum circuits.  
	As shown in Fig.~\ref{fig:bagging_workflow}(a), each quantum regressor consists of three parts: (i) quantum feature encoding, (ii) parameterized variational circuit, and (iii) classical linear readout. Here, we discuss about these three parts in details.\\
	
\noindent(i) \textbf{Quantum feature encoding}. Initially, all \(N_q\) qubits are prepared in the state \(\lvert 0\rangle\), giving the initial quantum state \(\lvert 0\rangle^{\otimes N_q}\). Each input feature is then encoded onto a qubit through a rotation about the \(y\) axis, with the rotation angle determined by the feature value. The encoding operation on qubit \(j\) is defined as
	\begin{equation}
		R_y(\pi\eta_j)
		=\exp\left(-i\frac{\pi\eta_j Y}{2}\right),
		\label{eq:ry_encoding}
	\end{equation}
	where $Y$ is the Pauli-$Y$ operator and
	$\eta_j{\in}\Phi(\mathbf{x})$ denotes the feature assigned to qubit $j$, selected from the feature pool defined in Eq.~\eqref{eq:feature_pool}. We consider two feature-to-qubit mapping strategies to investigate how different encoding schemes influence the nonlinear expressive capacity of quantum circuits and their regression performance. In the first strategy, without repetition, $N_q$ distinct features are randomly selected from the feature pool $\Phi(\mathbf{x})$ to construct a feature subset $\phi(\mathbf{x})$, and each selected feature is individually encoded onto one of the $N_q$ qubits. In the second strategy, with repetition, $N_q/2$ distinct features are randomly selected for an even number of qubits, or $(N_q{-}1)/2$ for an odd number of qubits. Each selected feature is then encoded onto a pair of adjacent qubits. When $N_q$ is odd, the last qubit receives no feature-encoding rotation and remains in the state $\lvert 0\rangle$ during feature encoding, but still participates in the subsequent variational circuit. Fig.~\ref{fig:bagging_workflow}(b) illustrates both mapping strategies using an eight-qubit circuit as an example. Repeating a feature across multiple qubits can enrich the nonlinear dependence of the circuit output on that feature, providing access to more complex functional relationships without increasing the polynomial order of the candidate feature pool \\

\noindent(ii) \textbf{Parameterized variational quantum circuit}. This part consists of \(L\) variational layers, each using a hardware-efficient ansatz. Each layer \(\ell\) applies a sequence of single-qubit rotations \(R_x(\theta_{\ell,j,1})\), \(R_z(\theta_{\ell,j,2})\), and \(R_x(\theta_{\ell,j,3})\) to every qubit \(j\), followed by a chain of nearest-neighbor CNOT gates to generate entanglement between the qubits, see Fig.~\ref{fig:bagging_workflow}(a). All trainable rotation angles are collectively denoted by \(\bm{\theta}{=}\{\theta_{\ell,j,r}\}\), where \(\ell{=}1,\ldots,L\), \(j{=}1,\ldots,N_q\), and \(r{=}1,2,3\), giving a total of \(3LN_q\) variational parameters. Together, these layers implement the parameterized unitary operation \(U(\bm{\theta})\), which acts on the feature-encoded quantum state.\\
	
	\noindent(iii) \textbf{Classical linear readout}. At the end of the circuit, the Pauli-$Z$ expectation value of each qubit is measured and passed to a trainable classical linear readout to obtain the prediction
	\begin{equation}
		\widehat{y}(\mathbf{x})
		=
		\sum_{j=1}^{N_{\mathrm{meas}}}
		c_j\langle Z_j\rangle,
		\label{eq:regressor_readout}
	\end{equation}
	where $\widehat{y}(\mathbf{x})$ is the prediction of $y(\mathbf{x})$ for sample \(\mathbf{x}\), $N_{\mathrm{meas}}$ is the number of
	measured qubits, $c_j$ is the trainable classical readout
	coefficient associated with qubit $j$, and
	$\langle Z_j\rangle$ denotes its Pauli-$Z$ expectation
	value. The readout coefficients \(c_j\) and variational circuit parameters \(\theta_{j,r}\), with \(r{=}1,2,3\), are jointly optimized using the Adam optimizer. \\

	For each regressor, the training procedure consists of the following six steps, with a total of 30 training epochs. The quantum circuit architecture described above and this training procedure provide a common framework for the regressors employed in the different ensemble learning approaches presented in the following sections.
	\begin{wideexplanationbox}{Training procedure
for an individual quantum regressor}

\begin{enumerate}[label=(\arabic*), leftmargin=*, itemsep=6pt]
		
		\item We select a subset of features from the candidate
		pool, $\phi(\mathbf{x}){\subset}\Phi(\mathbf{x})$,
		with the number of selected features determined by
		the feature-to-qubit mapping strategy.
		For example, an 8-qubit circuit ($N_q{=}8$)
		with repetition can use the feature subset
		$\phi(\mathbf{x}){=}
		\{x_1,x_1x_3,x_3^2,x_4\}$. 
		
		\item Before starting the first epoch, the training
		samples, $\{\mathbf{x}_i\}_{i=1}^{640}$,
		are randomly shuffled and divided into 10
		mini-batches, each containing $N_{\mathrm{batch}}{=}64$ different samples.
		This feature subset remains fixed throughout
		the training of the regressor. The variational circuit parameters are randomly
		initialized, and the classical readout coefficients
		are initially set to one.
		
		\item Starting with the first mini-batch, containing
		samples $\mathbf{x}_1,\ldots,\mathbf{x}_{64}$,
		we calculate the corresponding values of the
		selected features $\phi(\mathbf{x}_i)$ for each
		sample and encode them into the quantum circuit.
		The regressor then generates 64 predictions,
		$\hat{y}_1,\ldots,\hat{y}_{64}$,
		based on Eq.~\eqref{eq:regressor_readout},
		using the same current variational circuit
		parameters and classical readout coefficients.
		
		\item The 64 predictions are compared with their
		corresponding target values to calculate the
		loss using the loss function defined for each
		approach. The Adam optimizer then uses the
		loss gradients to update all circuit
		parameters and classical readout coefficients for each batch. 
		
		\item We repeat steps (3) and (4) for the remaining
		mini-batches, using the updated trainable
		parameters for each subsequent mini-batch,
		until all 640 training samples have been
		processed. Therefore, in each epoch the parameters are updated ten times, one time for each batch. This completes one epoch.
		
		\item We proceed to the next epoch, and repeat steps (3)--(5),
		starting from the trainable parameters obtained
		at the end of the previous epoch.
		After 30 epochs, we obtain the trained
		variational circuit parameters and classical
		readout coefficients, which are used for
		test-set evaluation.\\
		
	\end{enumerate}

\end{wideexplanationbox}
	After finishing the training, the regression performance is quantified on the test set using the coefficient of determination $R^2$, defined as
	\begin{equation}
		R^2 =
		1-
		\frac{\sum_{i=1}^{N_{\mathrm{test}}}\left[y(\mathbf{x}_i)-\widehat{y}(\mathbf{x}_i)\right]^2}
		{\sum_{i=1}^{N_{\mathrm{test}}}
        \left[y(\mathbf{x}_i)-\overline{y}\right]^2},
		\label{eq:r2}
	\end{equation}
	where $N_{\mathrm{test}}$ is the number of test samples, $y(\mathbf{x}_i)$ and $\widehat{y}(\mathbf{x}_i)$ are the true and predicted target values, respectively, and $\overline{y}{=}\frac{1}{N_{\mathrm{test}}}\sum_{i=1}^{N_{\mathrm{test}}}y(\mathbf{x}_i)$ is the mean of the true target values in the test set. A value of $R^2{=}1$ indicates perfect prediction, $R^2{=}0$ means the regression model has the same total squared prediction error as a simple model that always predicts the mean of the test targets, and $R^2{<}0$ indicates worse performance than this baseline. 
	
	\section{Bagging Ensemble Learning}
	\label{sec:bagging}

	\begin{figure*}[t]
		\centering
		
		\includegraphics[
		width=0.5\textwidth,
		height=0.4\textheight,
		keepaspectratio
		]{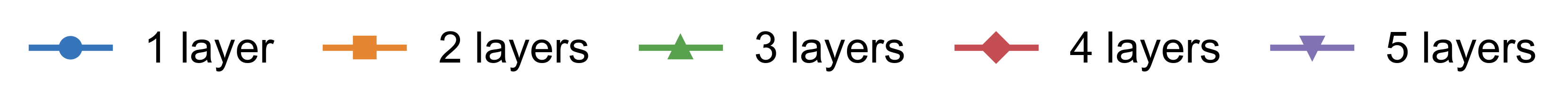}
		
		\vspace{1mm}
		
		\begin{minipage}{0.72\textwidth}
			\centering
			
			\begin{subfigure}{0.495\textwidth}
				\begin{overpic}[width=\textwidth]
					{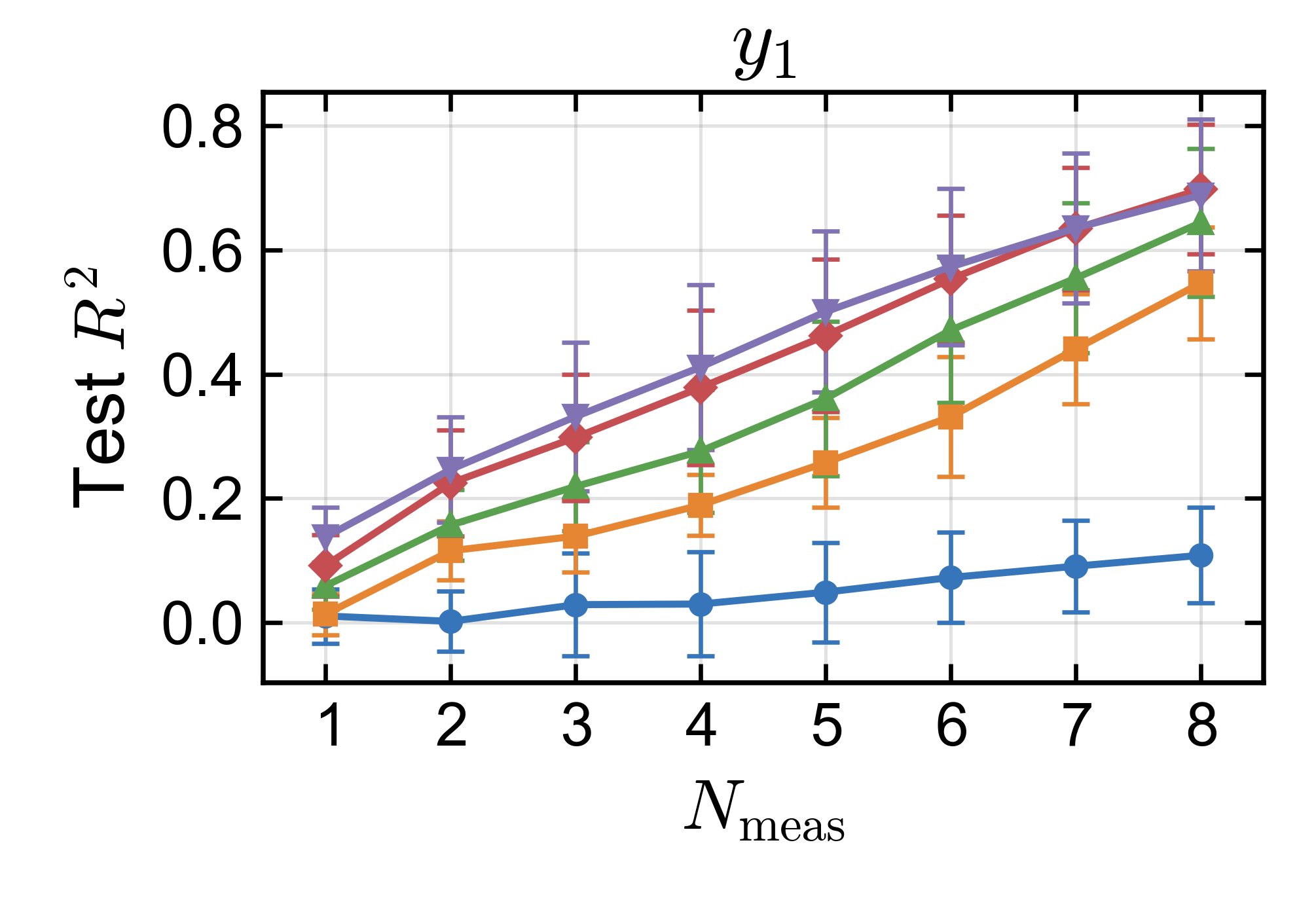}
					\put(2,60){{(a)}}
				\end{overpic}
				\label{fig:mult2}
			\end{subfigure}%
			\hfill%
			\begin{subfigure}{0.495\textwidth}
				\begin{overpic}[width=\textwidth]
					{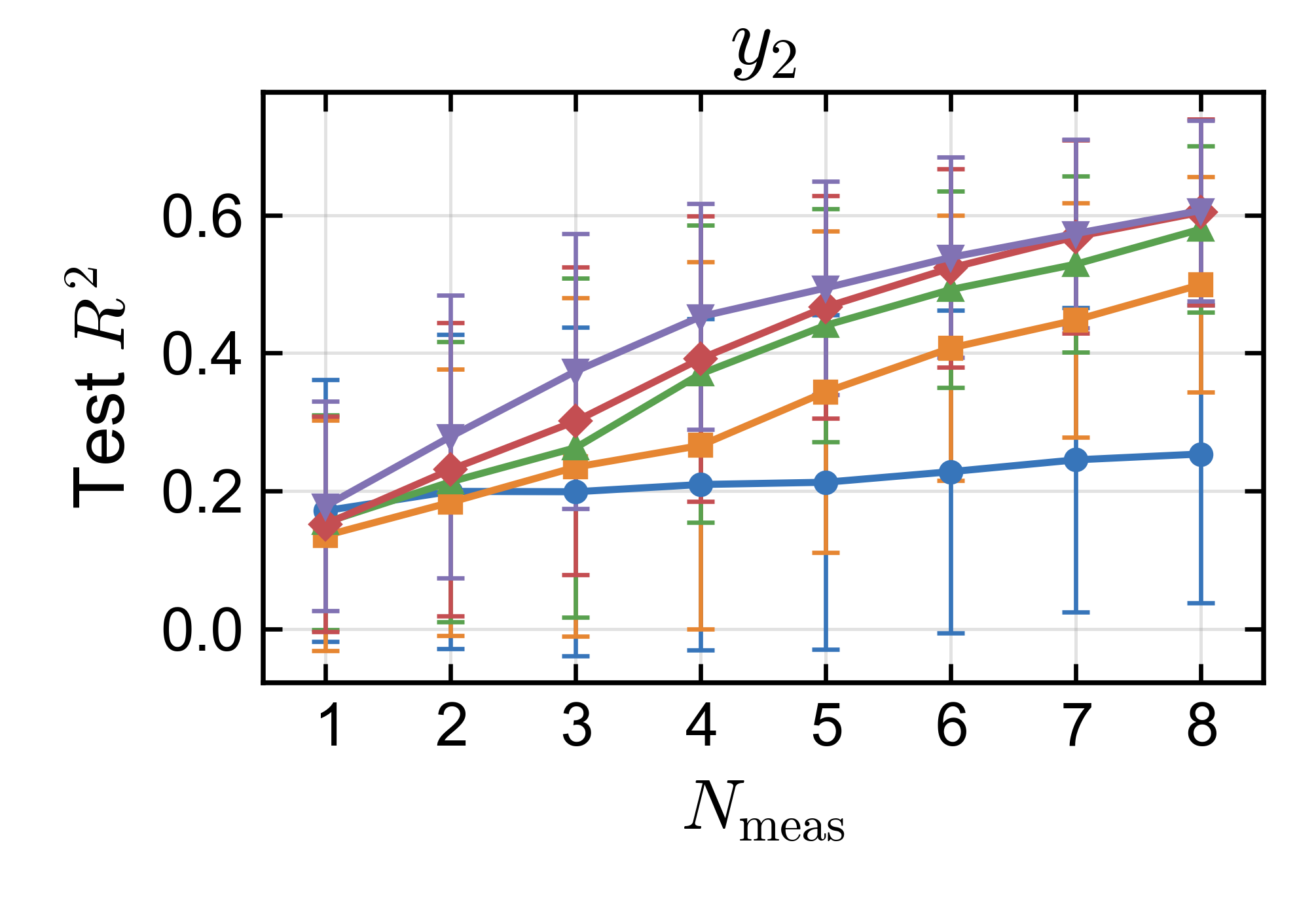}
					\put(2,60){{(b)}}
				\end{overpic}
				\label{fig:mult4}
			\end{subfigure}
			
			\vspace{-7.0mm}
			
			\begin{subfigure}{0.495\textwidth}
				\begin{overpic}[width=\textwidth]
					{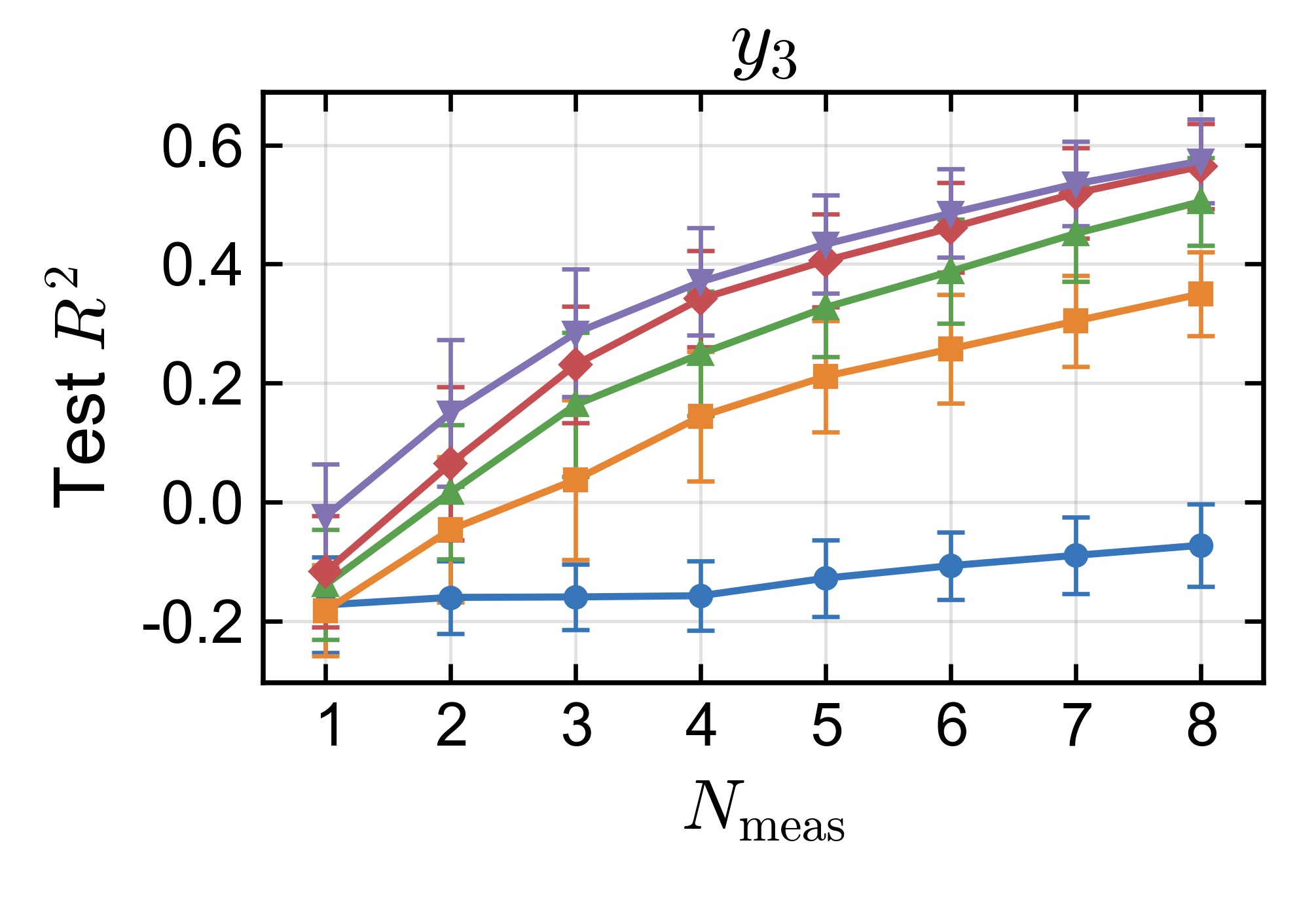}
					\put(2,60){{(c)}}
				\end{overpic}
				\label{fig:mult5}
			\end{subfigure}%
			\hfill%
			\begin{subfigure}{0.495\textwidth}
				\begin{overpic}[width=\textwidth]
					{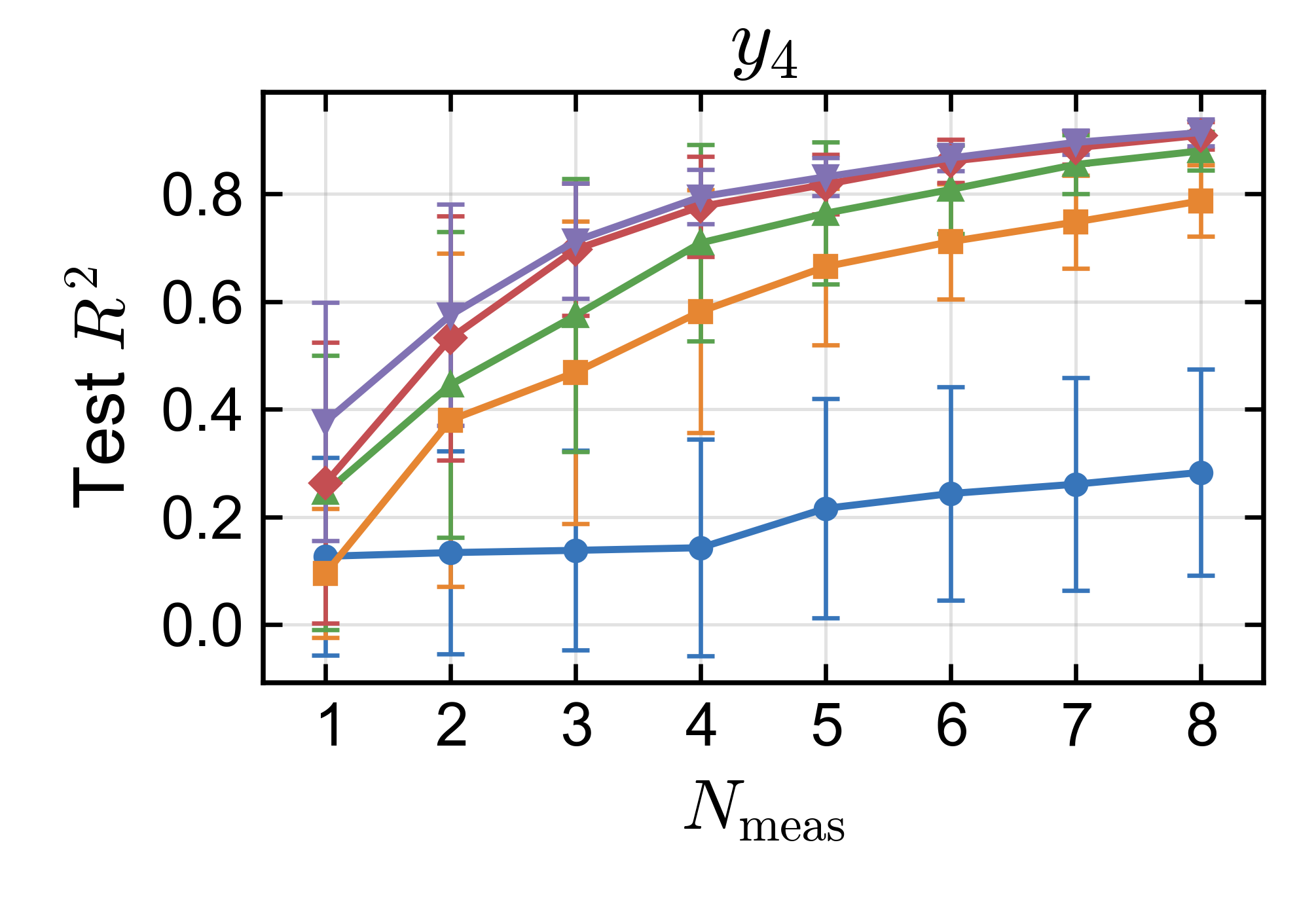}
					\put(2,60){{(d)}}
				\end{overpic}
				\label{fig:mult6}
			\end{subfigure}
		\end{minipage}
		\vspace{-0.5cm}
		\caption{Test-set $R^2$ of the Bagging approach as a function of the number of measured qubits $N_{\mathrm{meas}}$ included in the readout. Panels (a)--(d) correspond to $y_1$--$y_4$, respectively. The curves compare different circuit depths, and the error ranges indicate the variation over ten independent runs. All results are obtained using feature encoding without repetition.}
		\label{fig:bagging_nmeas}
	\end{figure*}

	We start our ensemble learning evaluation with bootstrap aggregating (which is commonly known as Bagging) approach. Bagging  is an ensemble-learning technique in which multiple models are trained independently and their predictions are combined to improve predictive performance and reduce model variance~\cite{breiman1996bagging,buhlmann2002analyzing}. For regression, the ensemble prediction is typically obtained by averaging the outputs of independently trained regressors. In this section, we investigate the Bagging approach as a baseline for evaluating ensemble learning techniques. \\
	
	The Bagging ensemble learning combines multiple independently trained regressors by averaging their predictions. For an ensemble of $B$ regressors, the procedure is as follows. Each regressor is trained independently on its corresponding bootstrap dataset using the six steps described in the previous section. The training loss of the \(b\)-th regressor is defined as
\begin{equation}
    \mathcal{L}_{\mathrm{train}}^{(b)}
    =
    \frac{1}{N_{\mathrm{batch}}}
    \sum_{i=1}^{N_{\mathrm{batch}}}
    \left[\widehat{y}^{(b)}(\mathbf{x}_i)-y(\mathbf{x}_i)\right]^2,
    \label{eq:bagging_mse}
\end{equation}
where \(b{=}1,\ldots,B\) labels the regressors in the Bagging ensemble.  Each regressor has independently initialized and optimized until all $B$ regressors are trained. After that, for each test sample $\mathbf{x}_i$, the $b$-th regressor generates a prediction $\widehat{y}^{(b)}(\mathbf{x}_i)$ using the classical readout defined in Eq.~\eqref{eq:regressor_readout}. The final prediction for each test sample is obtained by averaging the predictions of the $B$ regressors as
	\begin{equation}
		\widehat{y}^{\mathrm{bag}}(\mathbf{x}_i)
		=
		\frac{1}{B}
		\sum_{b=1}^{B}
		\widehat{y}^{(b)}(\mathbf{x}_i).
		\label{eq:bagging_prediction}
	\end{equation}
	The above estimation comes from an ensemble of $B$ independent regressors and therefore can compensate the lack of accuracy in each individual trained regressor. \\

	\begin{figure*}[t]
		\centering
		\includegraphics[width=0.99\textwidth]{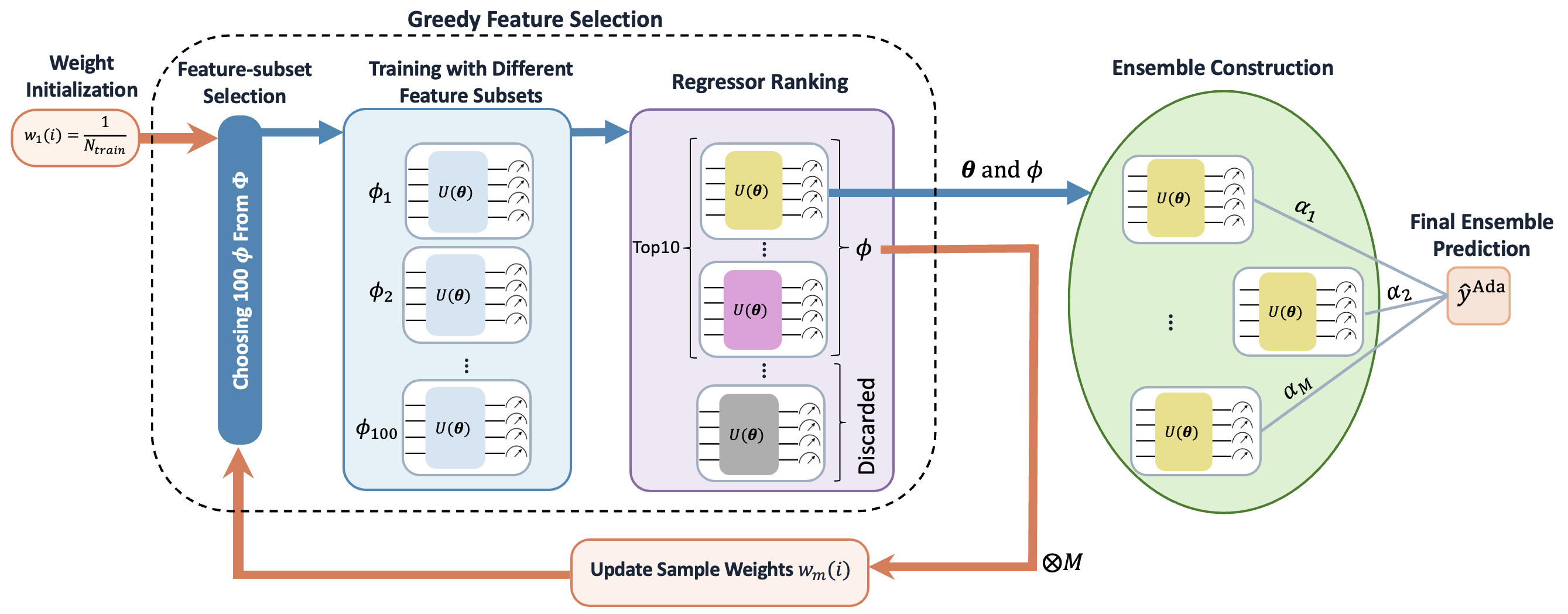}
		\caption{Schematic of the greedy AdaBoost algorithm. In each boosting round, 100 feature subsets, \(\), are selected from the candidate pool $\Phi(\mathbf{x})$, and a quantum regressor is trained for each subset. The trained candidates are ranked according to their prediction performance, and the best-performing regressor is selected. The ten highest-ranked feature subsets are retained for the next round, while the remaining candidates are discarded. The sample weights are then updated, and the procedure is repeated. Finally, the selected best regressors are combined through weighted averaging to obtain the ensemble prediction $\hat{y}^\mathrm{Ada}(\mathbf{x})$. }
		\label{fig:adaboost_workflow}
	\end{figure*}

To evaluate the regression performance of the Bagging approach, the evaluation is performed over ten independent runs on an 8-qubit circuit, considering different numbers of measured qubits, $N_{\mathrm{meas}}{\in}\{1,\ldots,8\}$, different circuit depths, and both feature-encoding strategies, with and without repetition. As shown in Fig.~\ref{fig:bagging_nmeas} for without repetition encoding and \(B{=}10\) regressors, the test \(R^2\) generally improves with increasing numbers of measured qubits, $N_{\mathrm{meas}}$, suggesting that additional measured qubits provide useful information for the regression task. This result will be used in the following section of this work and all qubits will be measured in each independent regressor. Furthermore, increasing the circuit depth also tends to improve predictive performance but after about \(L{=}3\) layers, the improvement starts to saturate. This saturation shows that increasing the circuit depth does not always enhance the accuracy, indicating the limitation of circuit expressivity in improving the regression performance. Therefore, to achieve further improvements better training strategies are required.\\

As illustrated in Fig.~\ref{fig:bagging_nmeas}, although the Bagging approach achieves reasonable $R^2$ values even with a five-layer circuit, its predictive performance still leaves considerable room for improvement, in particular in the case of $y_2$ and $y_3$. They therefore motivate the development of a better ensemble strategy, which we investigate through the greedy AdaBoost framework in the next section.

	\section{Greedy AdaBoost Ensemble Learning}
	\label{sec:adaboost}

	In this section, we investigate another ensemble-learning approach
	that adaptively guides the training of successive quantum regressors. Adaptive boosting (commonly known as AdaBoost) is an 
    ensemble-learning algorithm that uses a sequence of weak learners to construct a more accurate predictive model~\cite{freund1997decision}. Unlike the Bagging approach discussed in Sec.~\ref{sec:bagging}, where regressors are trained independently and their predictions are combined with equal weights, AdaBoost trains multiple regressors sequentially and assigns larger ensemble weights to regressors with smaller prediction errors. At the same time, it assigns greater importance to training samples with larger prediction errors, encouraging subsequent regressors to focus on samples that are more difficult to predict. In traditional AdaBoost, feature-subset selection is not an inherent part of the algorithm. However, the limited number of available qubits in the NISQ era restricts the number of input features that can be encoded into quantum circuits. To address this constraint, we incorporate an adaptive feature-selection procedure based on greedy algorithm~\cite{davis1997adaptive} into AdaBoost that identifies a suitable subset of features for the regressors at every boosting round. \\

The proposed greedy AdaBoost protocol is illustrated schematically in
Fig.~\ref{fig:adaboost_workflow}. The greedy algorithm selects and ranks
candidate feature subsets, while AdaBoost updates the training-sample
weights and determines the contribution of each selected regressor to
the final prediction. The complete procedure is divided into three stages: Stage 1 involves greedy feature selection, Stage 2 performs AdaBoost weight updates, and Stage 3 constructs the final ensemble. In the first boosting round (\(m{=}1\)), all training samples have equal weights, \(w_1(i){=}1/N_{\mathrm{train}}\), where \(i{=}1,\ldots,N_{\mathrm{train}}\).\\

\noindent\textbf{Stage 1: Greedy feature selection.}
This stage consists of three steps:

\begin{enumerate}[label=(\arabic*)]

\item \textbf{Feature-subset selection.}
Each quantum regressor uses a feature subset \(\phi\) selected from the candidate pool \(\Phi\). There are \(\binom{|\Phi|}{|\phi|}\) possible subsets, where \(|\Phi|\) and \(|\phi|\) denote the numbers of features in the pool and selected subset, respectively. Rather than evaluating all combinations, the algorithm randomly selects 100 feature subsets, \(\phi_k\) (with $k{=}1,\ldots,100$). In subsequent rounds, the ten highest-ranked feature subsets from the previous round are combined with 90 newly sampled subsets to form 100 candidates.

\item \textbf{Training with different feature subsets.}
Each selected feature subset is encoded into a quantum regressor. All 100 regressors are then independently trained using \(N_{\mathrm{train}}{=}640\) training samples, following the six-step procedure presented in the box in Sec.~\ref{sec:regressor}. During the training we use the sample weights determined by the AdaBoost procedure. This produces 100 trained regressors, \(\hat{y}^{(m,k)}(\mathbf{x})\) (with $k{=}1,\ldots,100$), where \(m\) denotes the boosting round and \(k\) labels the candidate regressor. 

In each round \(m\), the candidate regressors are trained by minimizing the weighted mean-squared error
\begin{equation}
\mathcal{L}_{\mathrm{train}}^{(m,k)}
=
\sum_{i=1}^{N_{\mathrm{batch}}}
w_m(i)
\left[\hat{y}^{(m,k)}(\mathbf{x}_i)-y(\mathbf{x}_i)\right]^2,
\label{eq:weighted_mse}
\end{equation}
where \(w_m(i)\) is the weight of training sample \(i\), and \(\hat{y}^{(m,k)}(\mathbf{x}_i)\) is the prediction of the \(k\)-th regressor in round \(m\). This loss gives samples with larger weights a greater influence on the training process.

\item \textbf{Regressor ranking.}
The 100 trained regressors are evaluated using \(N_{\mathrm{val}}{=}160\) validation samples and ranked according to their coefficient of determination, \(R^2\). The highest-ranked candidate is selected as the best regressor of round \(m\), denoted by \(\hat{y}^{(m,\mathrm{best})}(\mathbf{x})\). Its feature subset and trained variational parameters are stored for the final ensemble prediction. Meanwhile, the ten highest-ranked feature subsets are kept for the next boosting round. All previously trained candidate regressors are discarded, except for the selected best regressor.

\end{enumerate}

	\begin{figure*}[t]
		\centering
		
		\includegraphics[
		width=0.3\textwidth,
		height=0.055\textheight,
		keepaspectratio
		]{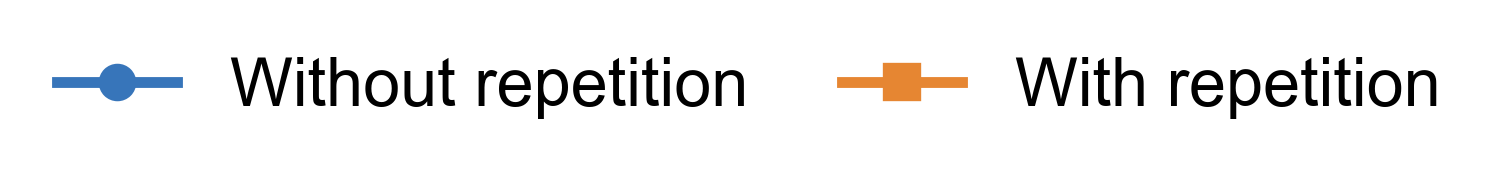}
		
		\vspace{0.5mm}
		
		\begin{subfigure}[c]{0.2435\textwidth}
			\begin{overpic}[percent,width=\linewidth]
				{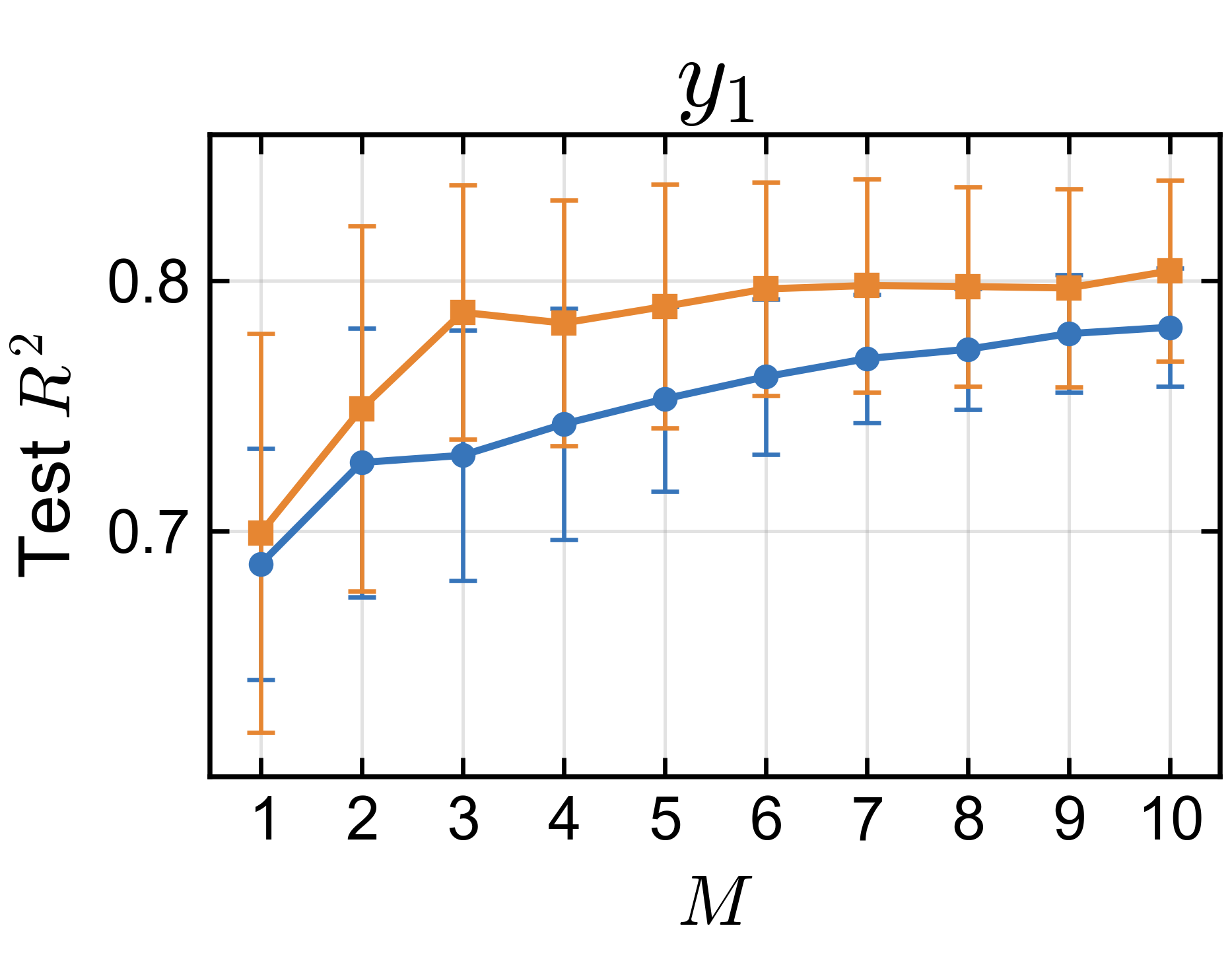}
				\put(4,75){(a)}
			\end{overpic}
			\label{fig:adaboost_learners_y1}
		\end{subfigure}%
		\hfill%
		\begin{subfigure}[c]{0.2435\textwidth}
			\begin{overpic}[percent,width=\linewidth]
				{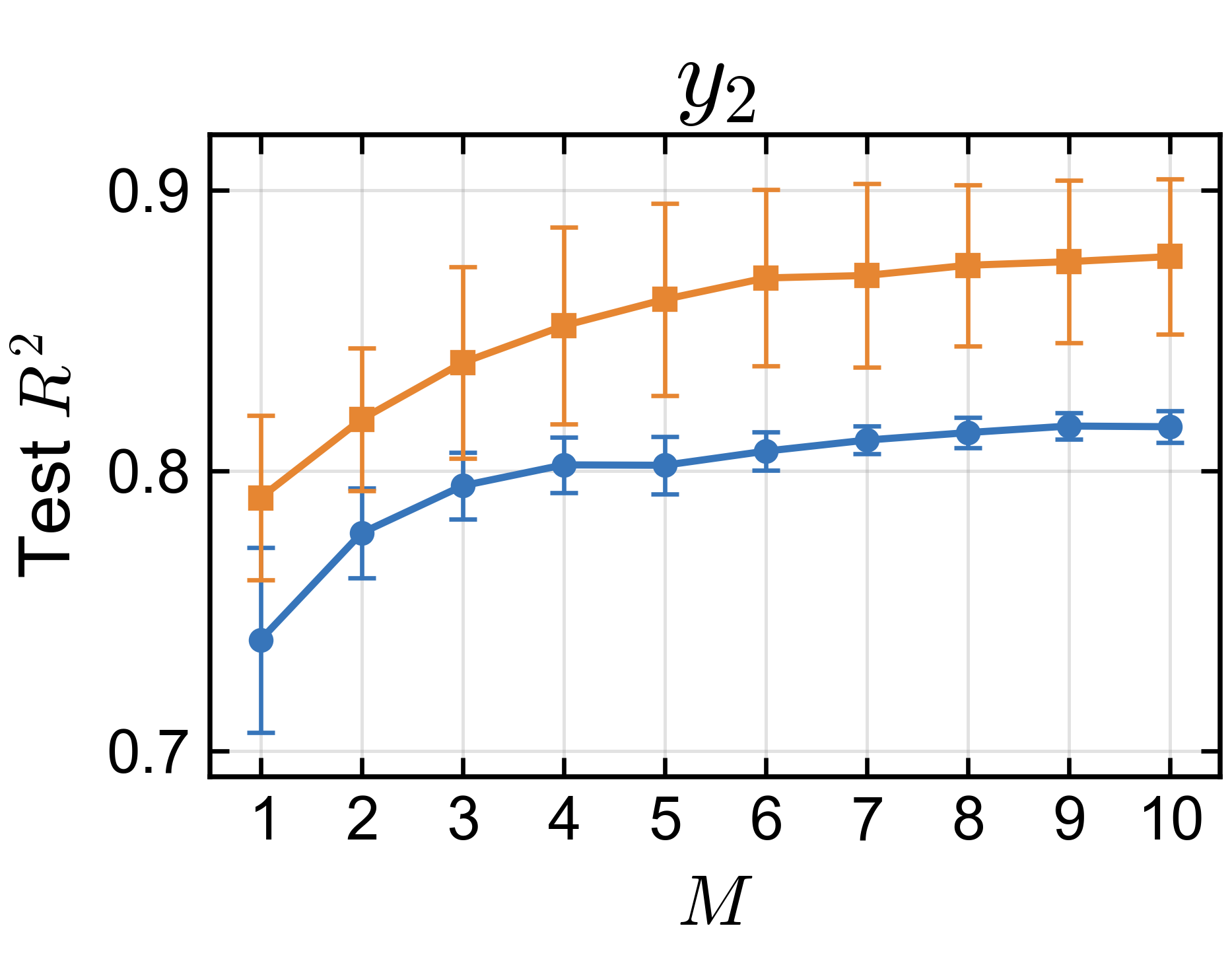}
				\put(1,75){(b)}
			\end{overpic}
			\label{fig:adaboost_learners_y2}
		\end{subfigure}%
		\hfill%
		\begin{subfigure}[c]{0.2435\textwidth}
			\begin{overpic}[percent,width=\linewidth]
				{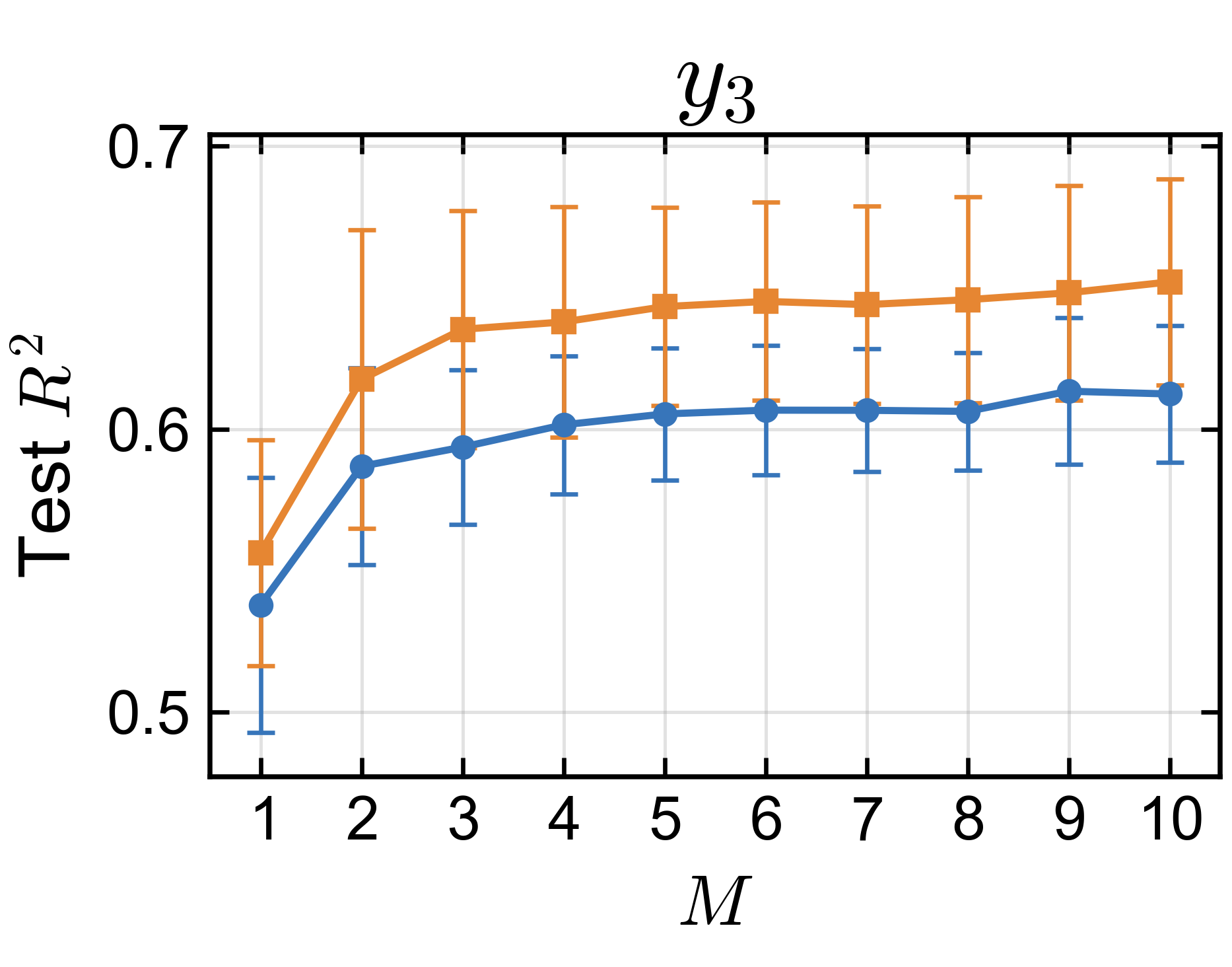}
				\put(1,75){(c)}
			\end{overpic}
			\label{fig:adaboost_learners_y3}
		\end{subfigure}%
		\hfill%
		\begin{subfigure}[c]{0.2435\textwidth}
			\begin{overpic}[percent,width=\linewidth]
				{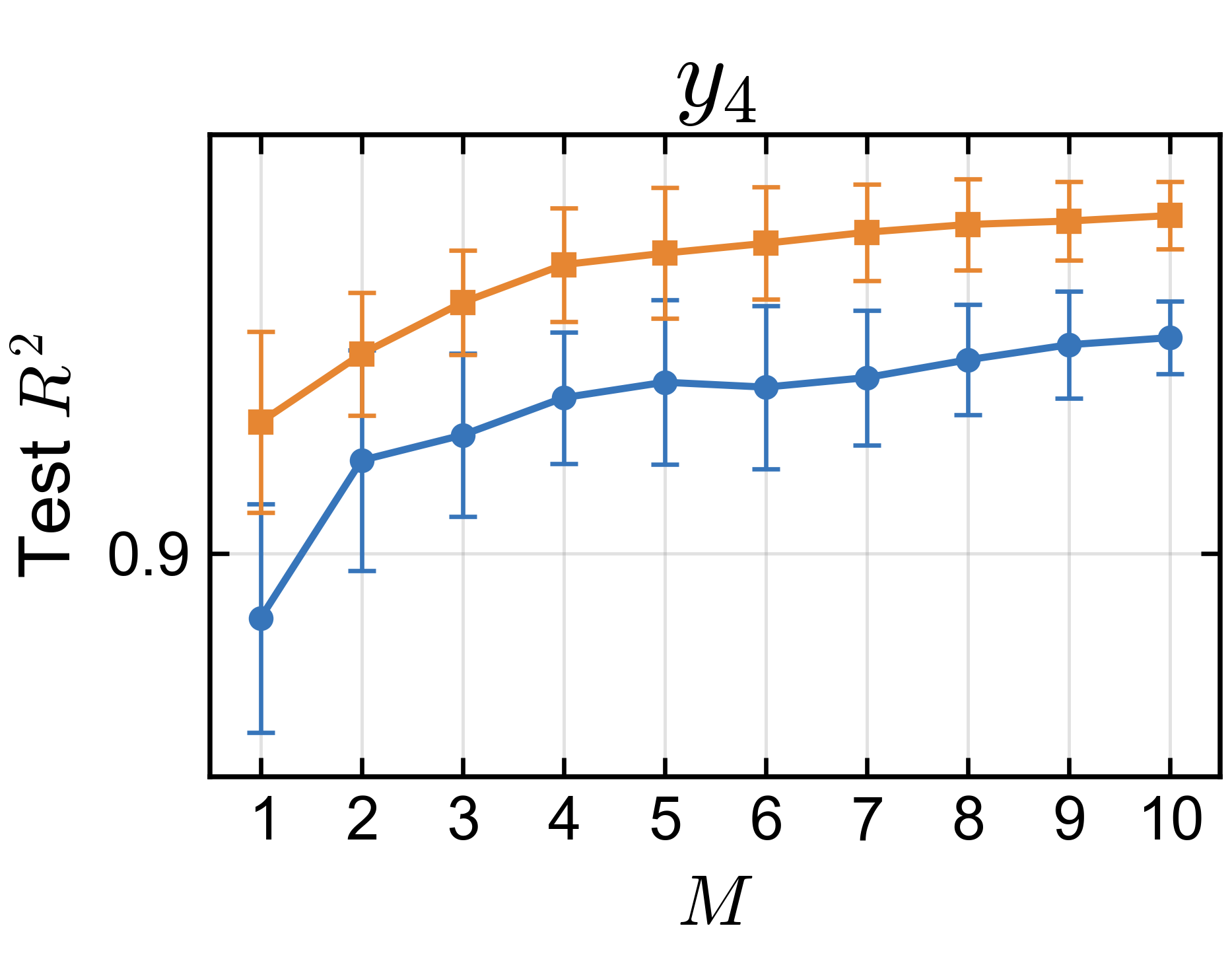}
				\put(5,75){(d)}
			\end{overpic}
			\label{fig:adaboost_learners_y4}
		\end{subfigure}
		\vspace{-10pt}
		\caption{Test-set $R^2$ of the AdaBoost ensemble as a function of the number of boosting rounds for an 8-qubit circuit ($N_q{=}8$), with and without repeated feature encoding. Panels (a)--(d) correspond to $y_1$--$y_4$, respectively. Error bars indicate the variation over ten independent runs.}
		\label{fig:adaboost_learners}
	\end{figure*}

\noindent\textbf{Stage 2: AdaBoost weight updates.}
After selecting the best regressor, AdaBoost calculates its prediction errors,
determines the boosting coefficients, and updates the training-sample weights.
This stage consists of three steps:

\begin{enumerate}[label=(\arabic*),start=4]

\item \textbf{Error calculation.}
After selecting the best regressor of round \(m\) based on validation \(R^2\), its absolute prediction error on each training sample is calculated as \(e^{(m,\mathrm{best})}(i){=}\lvert \hat{y}^{(m,\mathrm{best})}(\mathbf{x}_i){-}y(\mathbf{x}_i)\rvert\). The normalized error is then defined as
\begin{equation}
    \tilde{e}^{(m,\mathrm{best})}(i)
    =
    1-\exp\left[-\frac{e^{(m,\mathrm{best})}(i)}{E_m^{\mathrm{th}}}\right],
    \label{eq:normalized_error}
\end{equation}
where \(E^\mathrm{th}_{m}\) is the error threshold that 95\% of training samples meet. This transformation (\(e^{(m,\mathrm{best})}(i){\rightarrow}\tilde{e}^{(m,\mathrm{best})}(i)\)) maps the absolute prediction errors to the interval \([0,1)\), such that \(0{\leq}\tilde{e}^{(m,\mathrm{best})}(i){<}1\), while reducing the influence of exceptionally large errors. The weighted normalized error of round \(m\) is calculated as
\begin{eqnarray}
E_m{=}\sum_{i=1}^{N_{\mathrm{train}}}w_m(i)\tilde{e}^{(m,\mathrm{best})}(i).
\end{eqnarray}

\item \textbf{Boosting coefficients.}
Two boosting coefficients are used in the AdaBoost procedure: \(\beta_m\), which controls the sample-weight updates, and \(\alpha_m\), which determines the contribution of the selected regressor to the final ensemble prediction. These coefficients are calculated as
\begin{equation}
    \beta_m =
    \frac{E_m}{1-E_m+\varepsilon},
    \qquad
    \alpha_m =
    \ln\left(\frac{1}{\beta_m+\varepsilon}\right),
    \label{eq:weak_weight}
\end{equation}
where \(\varepsilon{=}10^{-10}\) is a numerical safeguard against division by zero.

\item \textbf{Update sample weights.}
Then, the sample weights for the next boosting round are updated according to
\begin{equation}
    w_{m+1}(i)
    =
    \frac{1}{Z_m}w_m(i)\beta_m^{1-\tilde{e}^{(m,\mathrm{best})}(i)},
    \label{eq:sample_weight_update}
\end{equation}
where \(Z_m{=}\sum_{i=1}^{N_{\mathrm{train}}}w_m(i)\beta_m^{1-\tilde{e}^{(m,\mathrm{best})}(i)}\) is a normalization factor to ensure the updated weights sum to one. Through this update, samples with larger normalized errors \(\tilde{e}^{(m,\mathrm{best})}(i)\) receive greater relative weights \(w_{m+1}(i)\), encouraging the next regressor to focus on samples that are more difficult to predict.

\end{enumerate}

\noindent\textbf{Stage 3: Ensemble construction.} Stages 1 and 2 are repeated until \(M\) successful boosting rounds have been completed. The best regressor from each round and its corresponding ensemble weight \(\alpha_m\) are stored. 

\begin{enumerate}[label=(\arabic*),start=7]

\item \textbf{Final ensemble prediction.}
The final ensemble prediction is obtained as the weighted average of the best regressors selected across \(M\) boosting rounds,
\begin{equation}
    \hat{y}^{\mathrm{Ada}}(\mathbf{x})
    =
    \frac{
    \sum_{m=1}^{M}
    \alpha_m\hat{y}^{(m,\mathrm{best})}(\mathbf{x})
    }{
    \sum_{m=1}^{M}\alpha_m
    },
    \label{eq:ensemble_prediction}
\end{equation}
Based on the above equation, a smaller weighted normalized error \(E_m\) of a round leads to a smaller \(\beta_m\) and a larger ensemble weight \(\alpha_m\) (see Eq.~\eqref{eq:weak_weight}), giving better-performing regressors a greater contribution to the final prediction.

\end{enumerate}

\begin{figure}[t]
		\centering
		
		\includegraphics[
		width=0.48\textwidth,
		height=0.055\textheight,
		keepaspectratio
		]{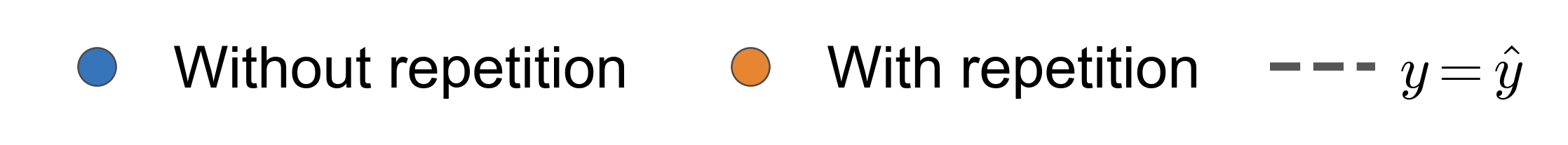}
		
		\vspace{0.5mm}
		
		\begin{subfigure}[c]{0.495\textwidth}
			\centering
			\begin{overpic}[percent,width=\linewidth]
				{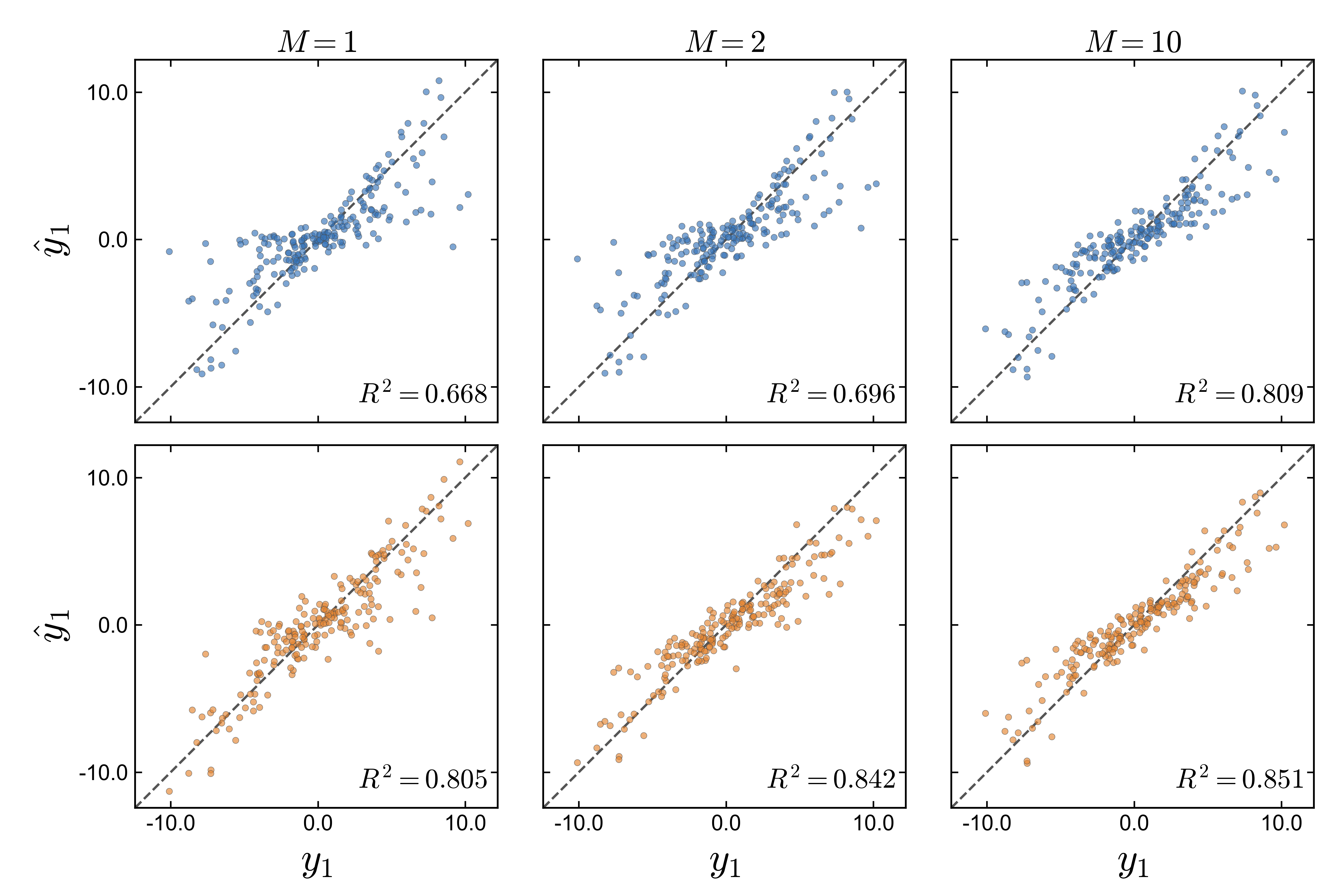}
				
				\put(11,60){\scriptsize{(a1)}}
				\put(42,60){\scriptsize{(a2)}}
				\put(73,60){\scriptsize{(a3)}}
				
				\put(11,31){\scriptsize{(a4)}}
				\put(42,31){\scriptsize{(a5)}}
				\put(73,31){\scriptsize{(a6)}}
			\end{overpic}
			\label{fig:adaboost_map1}
		\end{subfigure}%
		\hfill%
		\begin{subfigure}[c]{0.495\textwidth}
			\centering
			\begin{overpic}[percent,width=\linewidth]
				{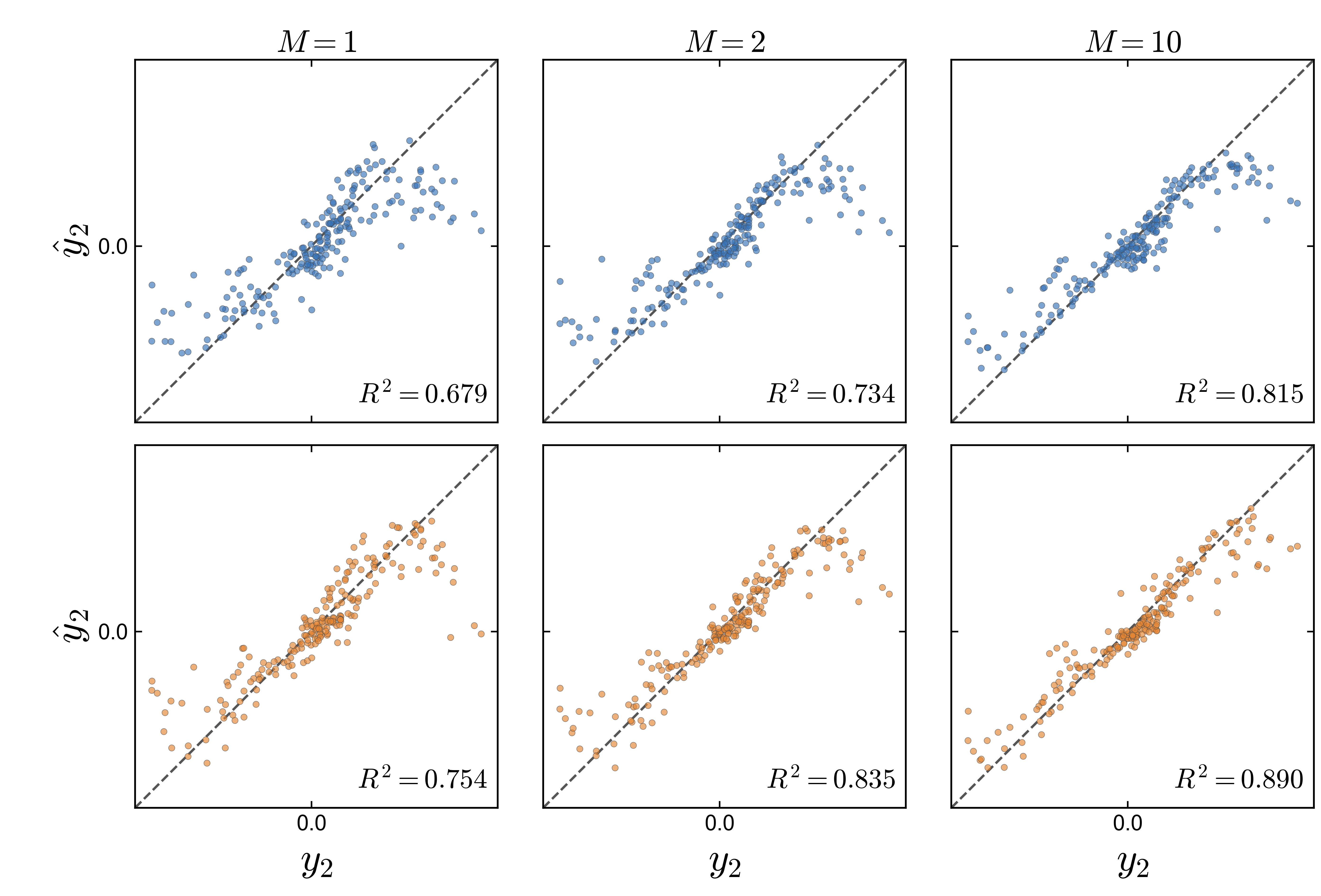}
				
				\put(11,60){\scriptsize{(b1)}}
				\put(42,60){\scriptsize{(b2)}}
				\put(73,60){\scriptsize{(b3)}}
				
				\put(11,31){\scriptsize{(b4)}}
				\put(42,31){\scriptsize{(b5)}}
				\put(73,31){\scriptsize{(b6)}}
			\end{overpic}
			\label{fig:adaboost_map2}
		\end{subfigure}
		
		\vspace{1mm}
		
		\begin{subfigure}[c]{0.495\textwidth}
			\centering
			\begin{overpic}[percent,width=\linewidth]
				{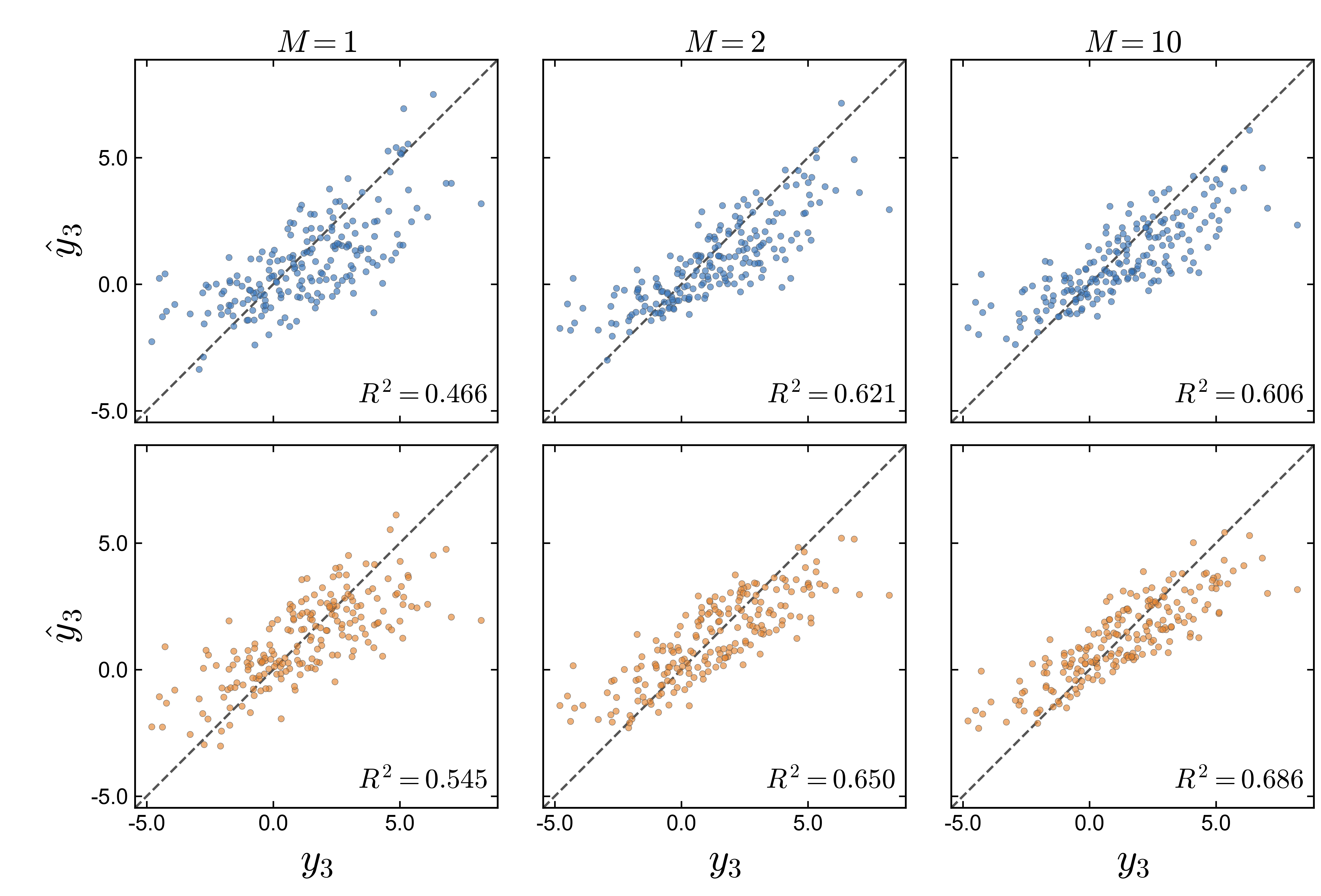}
				
				\put(11,60){\scriptsize{(c1)}}
				\put(42,60){\scriptsize{(c2)}}
				\put(73,60){\scriptsize{(c3)}}
				
				\put(11,31){\scriptsize{(c4)}}
				\put(42,31){\scriptsize{(c5)}}
				\put(73,31){\scriptsize{(c6)}}
			\end{overpic}
			\label{fig:adaboost_map3}
		\end{subfigure}%
		\hfill%
		\begin{subfigure}[c]{0.495\textwidth}
			\centering
			\begin{overpic}[percent,width=\linewidth]
				{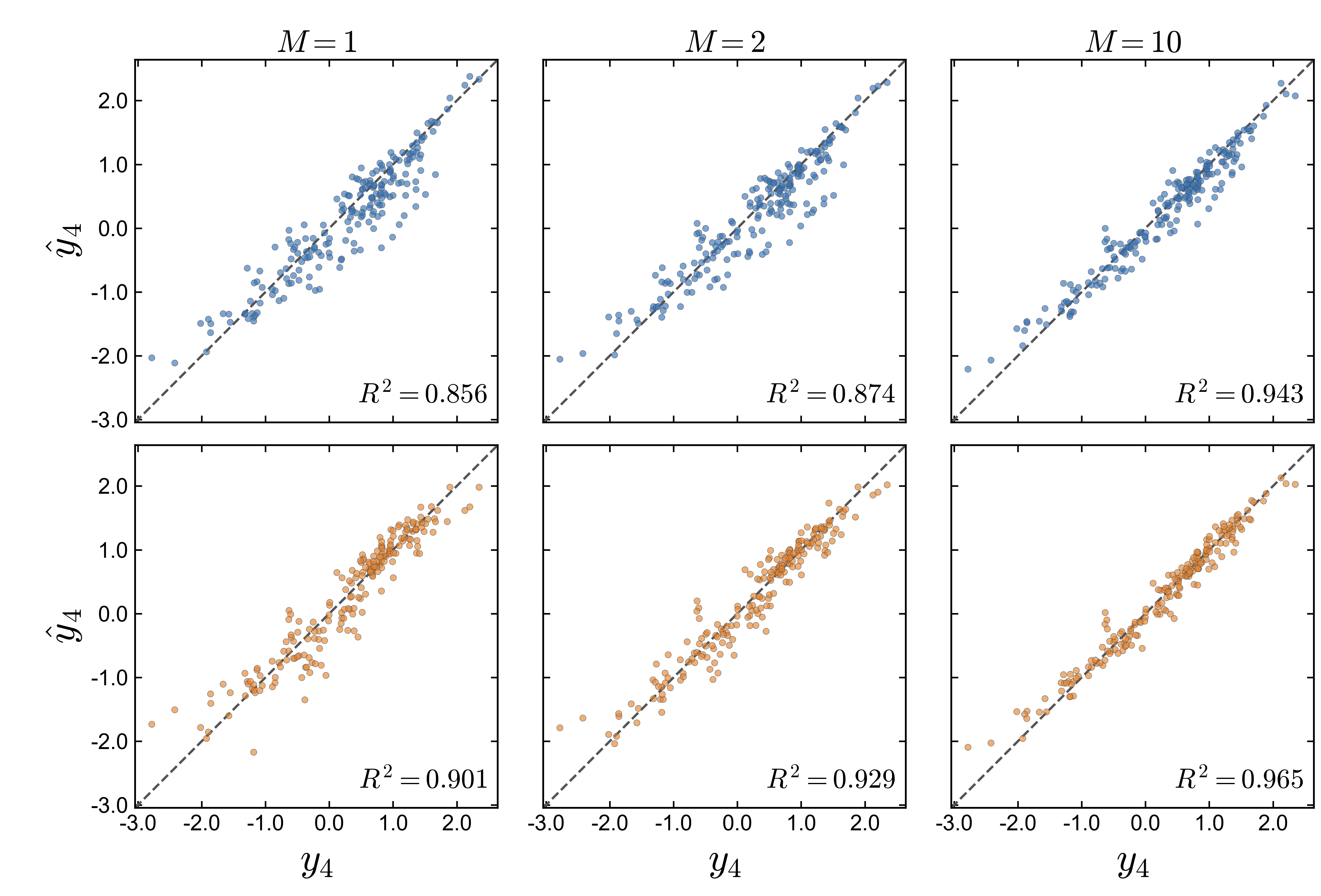}
				
				\put(11,60){\scriptsize{(d1)}}
				\put(42,60){\scriptsize{(d2)}}
				\put(73,60){\scriptsize{(d3)}}
				
				\put(11,31){\scriptsize{(d4)}}
				\put(42,31){\scriptsize{(d5)}}
				\put(73,31){\scriptsize{(d6)}}
			\end{overpic}
			\label{fig:adaboost_map4}
		\end{subfigure}
		
\caption{Predicted versus true test-set values obtained using the AdaBoost ensemble with an 8-qubit circuit ($N_q{=}8$) for four target functions. Panels (a1)--(a6), (b1)--(b6), (c1)--(c6), and (d1)--(d6) correspond to $y_1$, $y_2$, $y_3$, and $y_4$, respectively. For each target function, the upper and lower rows show the results without and with repeated feature encoding, respectively. The three columns correspond to $M{=}1$, $2$, and $10$ boosting rounds. The dashed lines indicate perfect predictions ($\widehat{y}{=}y$), and the test $R^2$ values are reported in each panel.}
		\label{fig:adaboost_scatter_4x4}
	\end{figure}

	This hybrid approach offers several advantages.
	First, the greedy feature-selection
	procedure allows different regressors to
	exploit complementary information from
	the feature pool without increasing the
	number of qubits of
	individual regressors.
	Second, adaptive sample weighting encourages
	successive regressors to focus on training
	samples that were poorly predicted in
	previous rounds.
	Finally, the error-dependent ensemble
	weights assign greater importance to
	more accurate regressors.
	Together, these mechanisms can improve
	predictive accuracy and stability compared
	with Bagging approach as we can see in the following simulation results.\\
	
	\begin{figure*}[t]
		\centering
		
		\includegraphics[
		width=0.3\textwidth,
		height=0.055\textheight,
		keepaspectratio
		]{figures/3_adaboost/legend_repetition.png}
		
		\vspace{0.5mm}
		
		\begin{subfigure}[c]{0.2435\textwidth}
			\begin{overpic}[percent,width=\linewidth]
				{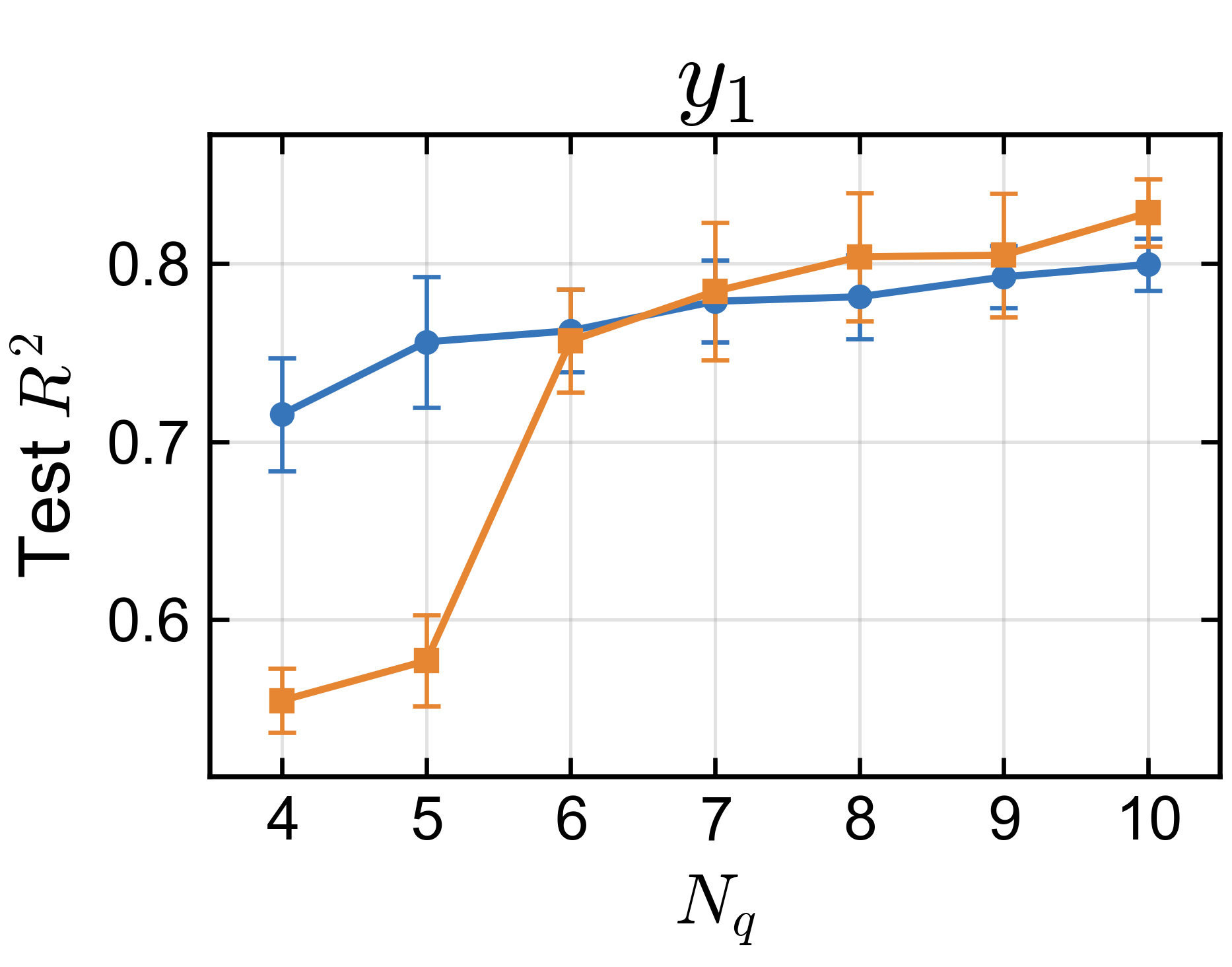}
				\put(4,75){(a)}
			\end{overpic}
			\label{fig:adaboost_qubits_y1}
		\end{subfigure}%
		\hfill%
		\begin{subfigure}[c]{0.2435\textwidth}
			\begin{overpic}[percent,width=\linewidth]
				{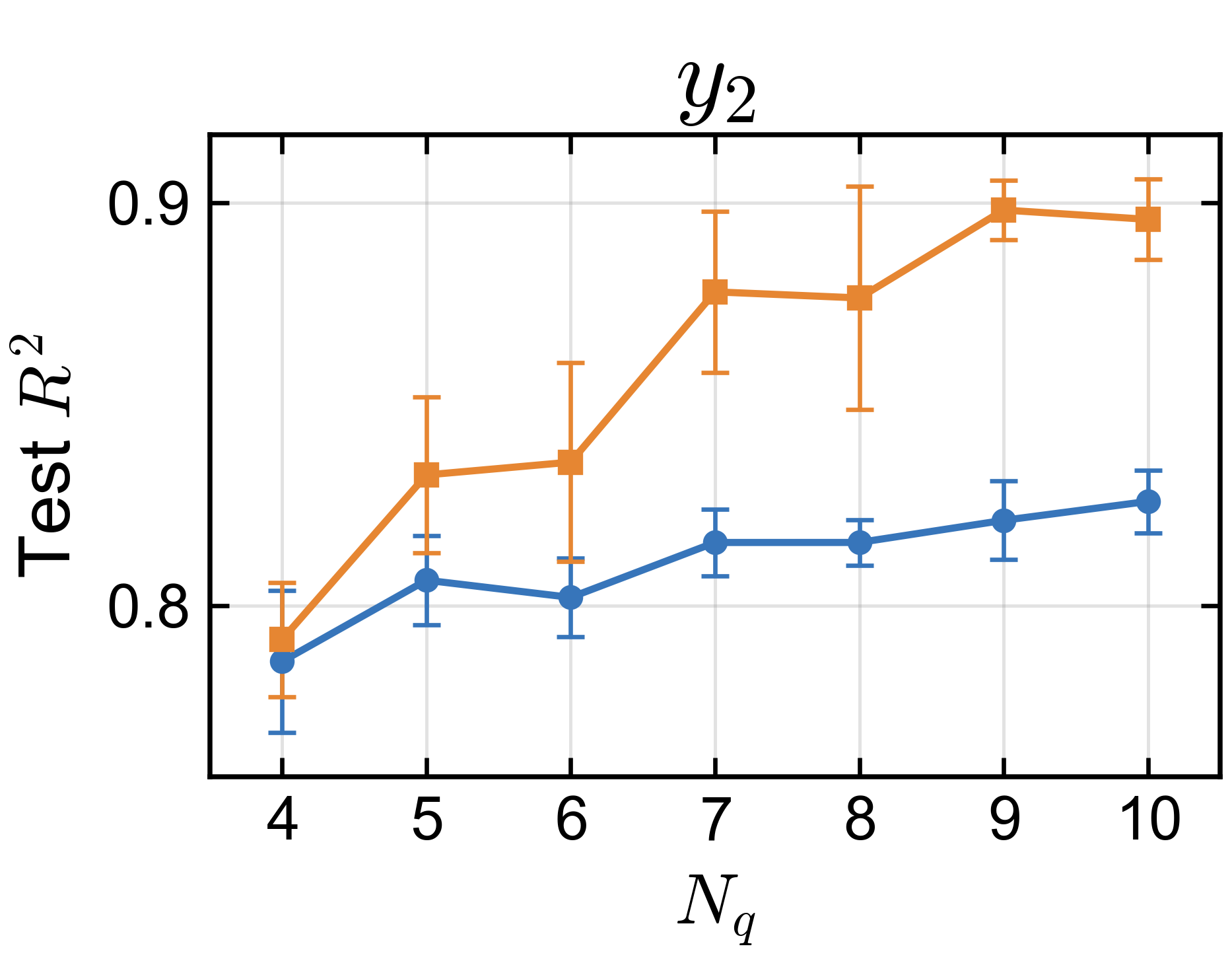}
				\put(1,75){(b)}
			\end{overpic}
			\label{fig:adaboost_qubits_y2}
		\end{subfigure}%
		\hfill%
		\begin{subfigure}[c]{0.2435\textwidth}
			\begin{overpic}[percent,width=\linewidth]
				{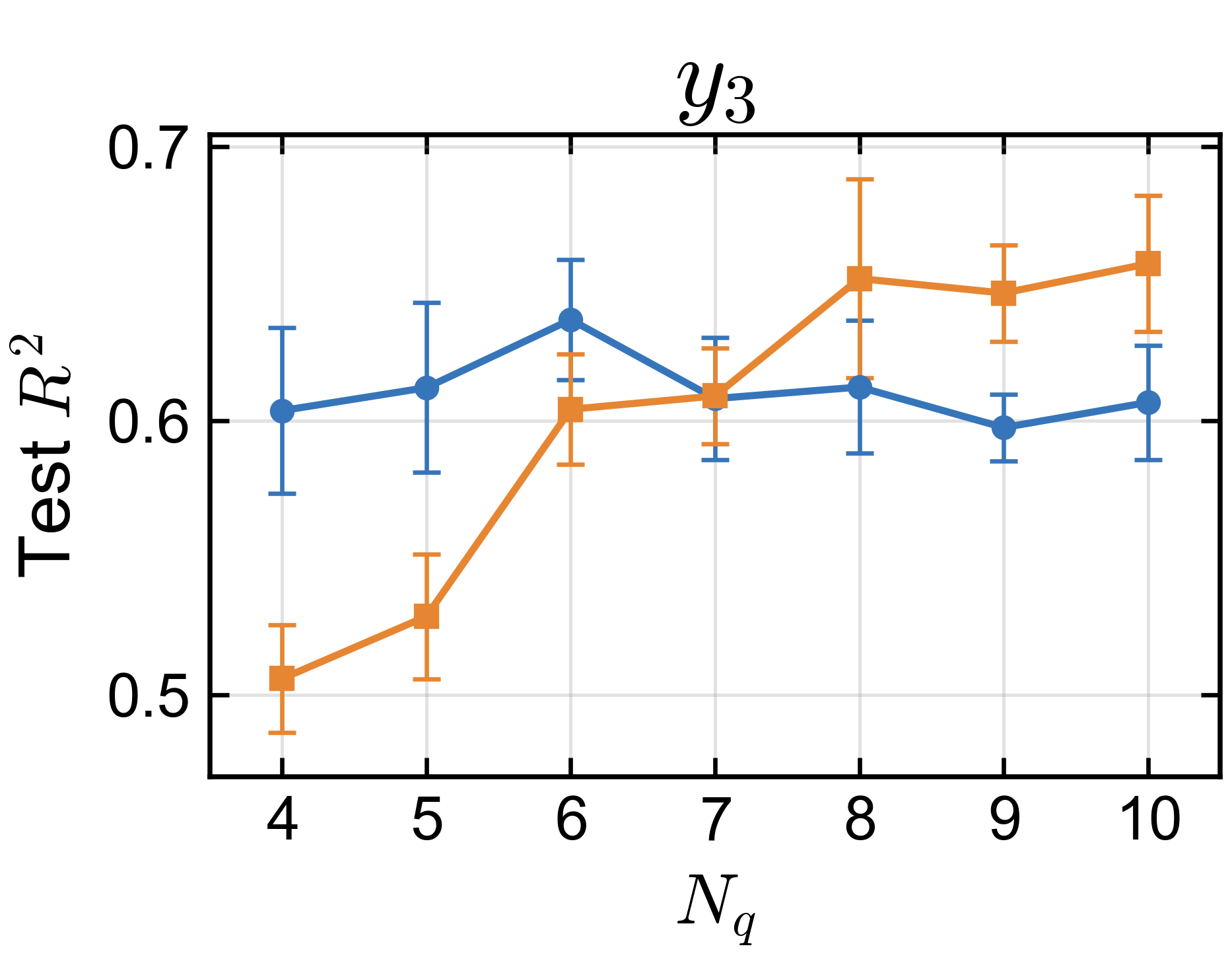}
				\put(1,75){(c)}
			\end{overpic}
			\label{fig:adaboost_qubits_y3}
		\end{subfigure}%
		\hfill%
		\begin{subfigure}[c]{0.2435\textwidth}
			\begin{overpic}[percent,width=\linewidth]
				{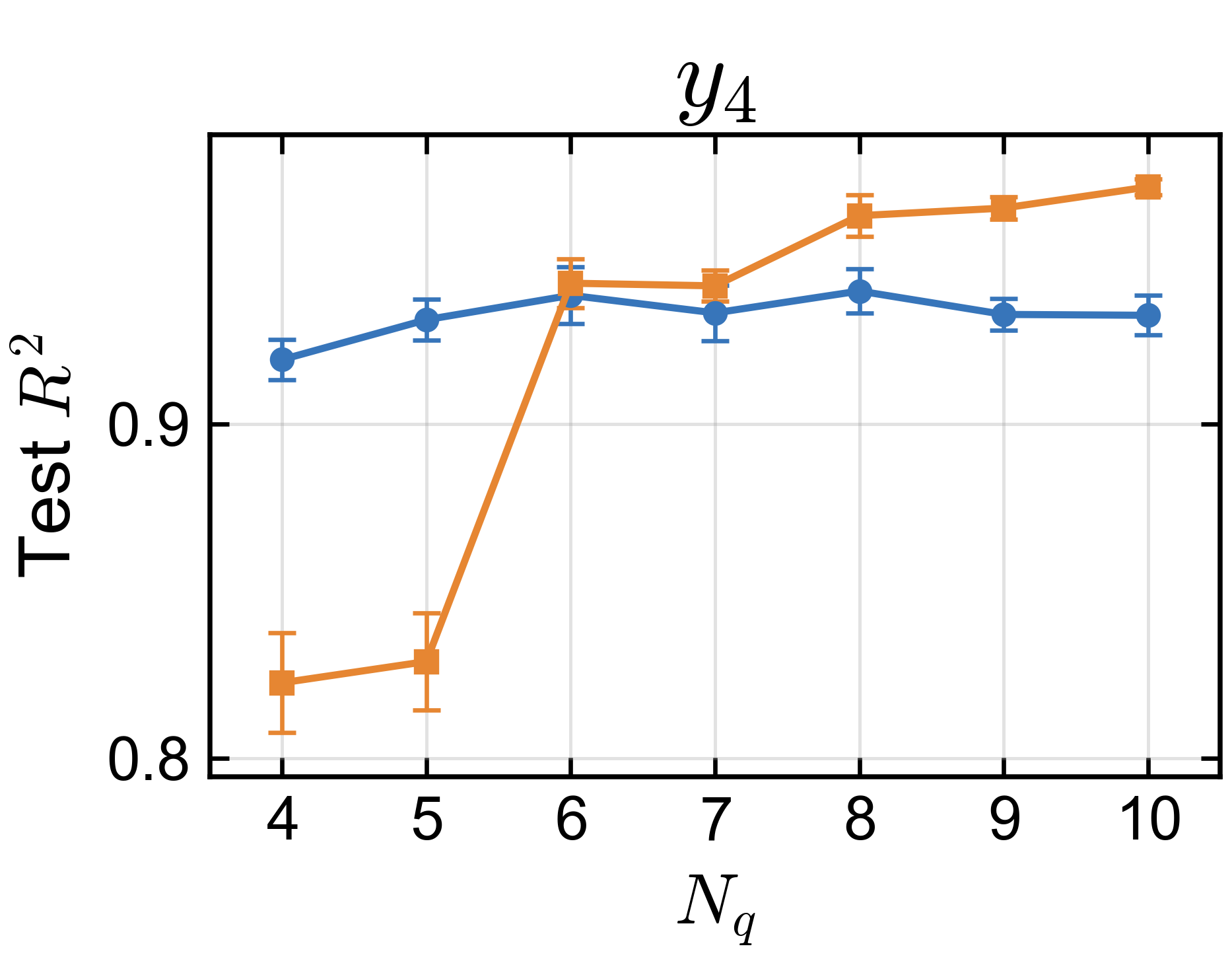}
				\put(5,75){(d)}
			\end{overpic}
			\label{fig:adaboost_qubits_y4}
		\end{subfigure}
		\vspace{-10pt}
\caption{
Test-set $R^2$ of AdaBoost as a function of the number of qubits for $M{=}10$ boosting rounds. Panels (a)--(d) correspond to $y_1$--$y_4$, respectively, and compare encoding with and without repetition.
}
		\label{fig:adaboost_qubits}
	\end{figure*}

In order to evaluate the performance of greedy AdaBoost, we first start with increasing the number of AdaBoost rounds, \(M\). As illustrated in Fig.~\ref{fig:adaboost_learners}, in the 8-qubit setting with only two layers, the test $R^2$ monotonically increases with the number of AdaBoost rounds across the four target mappings. This trend is happening for both with and without repetition encodings, though the repetition encoding strategy achieves significantly higher accuracy. A comparison of Figs.~\ref{fig:adaboost_learners} and~\ref{fig:bagging_nmeas} reveals that AdaBoost achieves higher regression performance using only $M{=}3$ rounds  with $L{=}2$ layers than the Bagging approach achieves using $B{=}10$ regressors with $L{=}5$ layers each. One might argue that AdaBoost evaluates 100 candidate regressors in each boosting round, whereas Bagging uses only ten regressors. However, Bagging simply averages the predictions of individual regressors without selecting the best-performing candidates, so the performance saturates quickly as the number of regressors increases.\\

To further illustrate the improvement with increasing ensemble size, Fig.~\ref{fig:adaboost_scatter_4x4} compares the predicted and true test values for all four target functions at different numbers of boosting rounds. As $M$ increases, the predictions generally become more concentrated around the ideal line $\widehat{y}{=}y$, indicating improved regression accuracy. However, there are some points which are not very well predicted. We find out that that those points are mostly sampled from the tails of the distributions, see Fig.~\ref{fig:target_distributions}. The reason that the regressors cannot predict those points is that they are rare events and not many of them appear in the training set. \\
	
We also investigate  the effect of qubit numbers on the accuracy of the AdaBoost algorithm. Therefore, we fix the number of AdaBoost rounds to \(M{=}10\) in which each regressor has $L{=}2$ layers. We increase the number of qubits from $N_q{=}4$ to $N_q{=}10$. As the number of qubits increases, both the number of features that can be encoded and the number of variational parameters change. The corresponding results are shown in Fig.~\ref{fig:adaboost_qubits}. At low qubit numbers, encoding without repetition provides broader coverage of distinct input features and achieves higher \(R^2\) values for several target mappings. When the number of qubits reaches eight to ten, repeated encoding attains a higher $R^2$. These results suggest that repeating features across qubits helps the circuit capture more complex nonlinear relationships, even when the candidate feature pool contains only first- and second-order polynomial terms. Consequently, despite using only first- and second-order input features, the proposed framework can achieve high prediction accuracy for target functions containing higher-order polynomial and fractional-power nonlinearities. Another important observation is that increasing the number of qubits generally improves regression accuracy, suggesting that by scaling up the quantum machines one can reach better regression performance. Furthermore, the results indicate that greedy AdaBoost can effectively exploit this increased capacity through appropriate feature selection and encoding strategies. \\

One of the important parameters for quantum regression evaluation is the circuit depth. Table~\ref{tab:r2_vs_layers} reports the AdaBoost test-set $R^2$ values for the four target mappings using an 8-qubit circuit ($N_q{=}8$) under both feature-encoding strategies, comparing a single regressor ($M{=}1$) with an ensemble of $M{=}10$ regressors. Overall, regression performance generally improves as the circuit depth increases, with particularly pronounced gains when moving from one to five variational layers. Across all configurations reported in Table~\ref{tab:r2_vs_layers}, the ensemble containing $M{=}10$ regressors achieves a higher mean $R^2$ than the corresponding single-regressor model with $M{=}1$.

	\begin{table*}[t]
		\centering
		
		\footnotesize
		\setlength{\tabcolsep}{3pt}
		\renewcommand{\arraystretch}{1.2}
		\setlength{\arrayrulewidth}{0.5pt}
		
		\resizebox{\textwidth}{!}{%
			\begin{tabular}{|c|c|*{5}{c|}|*{5}{c|}}
				\hline
				
				\multirow{2}{*}{Target}
				& \multirow{2}{*}{$M$}
				& \multicolumn{5}{c||}{\textbf{Without repetition}}
				& \multicolumn{5}{c|}{\textbf{With repetition}} \\
				\cline{3-12}
				
				& & $L=1$ & $L=2$ & $L=3$ & $L=4$ & $L=5$
				& $L=1$ & $L=2$ & $L=3$ & $L=4$ & $L=5$ \\
				\hline
				
				\multirow{2}{*}{$y_1$}
				& 1
				& $0.36{\pm}0.02$
				& $0.69{\pm}0.05$
				& $0.75{\pm}0.05$
				& $0.79{\pm}0.04$
				& $0.78{\pm}0.04$
				& $0.40{\pm}0.03$
				& $0.70{\pm}0.08$
				& $0.74{\pm}0.07$
				& $0.74{\pm}0.06$
				& $0.78{\pm}0.02$ \\
				
				& 10
				& $0.41{\pm}0.02$
				& $0.78{\pm}0.02$
				& $0.83{\pm}0.02$
				& $0.84{\pm}0.01$
				& $0.86{\pm}0.01$
				& $0.49{\pm}0.02$
				& $0.80{\pm}0.04$
				& $0.82{\pm}0.02$
				& $0.83{\pm}0.03$
				& $0.83{\pm}0.02$ \\
				\hline
				
				\multirow{2}{*}{$y_2$}
				& 1
				& $0.52{\pm}0.02$
				& $0.74{\pm}0.03$
				& $0.82{\pm}0.03$
				& $0.86{\pm}0.02$
				& $0.88{\pm}0.02$
				& $0.55{\pm}0.02$
				& $0.79{\pm}0.03$
				& $0.86{\pm}0.05$
				& $0.86{\pm}0.05$
				& $0.86{\pm}0.05$ \\
				
				& 10
				& $0.54{\pm}0.01$
				& $0.82{\pm}0.01$
				& $0.87{\pm}0.01$
				& $0.90{\pm}0.01$
				& $0.91{\pm}0.00$
				& $0.58{\pm}0.01$
				& $0.88{\pm}0.03$
				& $0.92{\pm}0.00$
				& $0.93{\pm}0.00$
				& $0.93{\pm}0.01$ \\
				\hline
				
				\multirow{2}{*}{$y_3$}
				& 1
				& $0.16{\pm}0.06$
				& $0.54{\pm}0.04$
				& $0.59{\pm}0.05$
				& $0.68{\pm}0.05$
				& $0.70{\pm}0.06$
				& $0.33{\pm}0.02$
				& $0.56{\pm}0.04$
				& $0.59{\pm}0.05$
				& $0.65{\pm}0.05$
				& $0.65{\pm}0.04$ \\
				
				& 10
				& $0.27{\pm}0.03$
				& $0.61{\pm}0.02$
				& $0.69{\pm}0.02$
				& $0.75{\pm}0.02$
				& $0.77{\pm}0.01$
				& $0.41{\pm}0.02$
				& $0.65{\pm}0.04$
				& $0.70{\pm}0.02$
				& $0.74{\pm}0.03$
				& $0.76{\pm}0.02$ \\
				\hline
				
				\multirow{2}{*}{$y_4$}
				& 1
				& $0.64{\pm}0.03$
				& $0.89{\pm}0.02$
				& $0.92{\pm}0.02$
				& $0.93{\pm}0.02$
				& $0.93{\pm}0.01$
				& $0.70{\pm}0.05$
				& $0.92{\pm}0.02$
				& $0.91{\pm}0.02$
				& $0.93{\pm}0.02$
				& $0.94{\pm}0.02$ \\
				
				& 10
				& $0.70{\pm}0.02$
				& $0.94{\pm}0.01$
				& $0.96{\pm}0.01$
				& $0.96{\pm}0.00$
				& $0.96{\pm}0.00$
				& $0.78{\pm}0.01$
				& $0.96{\pm}0.01$
				& $0.96{\pm}0.00$
				& $0.97{\pm}0.00$
				& $0.97{\pm}0.00$ \\
				\hline
				
			\end{tabular}%
		}

		\caption{
    Test-set $R^2$ of the AdaBoost ensemble as a function
    of circuit depth $L$ for four target mappings using
    an 8-qubit circuit ($N_q{=}8$).
    Results are reported for $M{=}1$ and $M{=}10$
    boosting rounds, with and without repeated feature encoding.
    Values represent the mean and standard deviation
    across ten independent runs.
}
		\label{tab:r2_vs_layers}
		
	\end{table*}

	\section{Conclusion} \label{sec:conc}

	In this work, we investigated the integration of feature selection and ensemble learning to enhance quantum regression performance, particularly for approximating highly nonlinear functions with limited quantum resources. We considered four benchmark functions with different mathematical structures, ranging from linear to fractional-power nonlinear mappings. A feature pool containing first- and second-order polynomial terms was constructed, and two feature-to-qubit mapping strategies, with and without repeated feature encoding, were investigated. We employed two ensemble learning approaches, namely Bagging and AdaBoost, with the latter incorporating a greedy feature-selection algorithm.\\
	
	Our numerical results demonstrate that adaptively using nonlinear features, and repetition can significantly improve regression performance by enhancing the quantum circuits' ability to learn complex nonlinear dynamics. Furthermore, the results also show that a feature pool containing only first- and second-order polynomial terms can be used to approximate functions with higher-order and fractional-power nonlinearities if the repetition is used. Our simulations show that greedy AdaBoost achieves higher prediction accuracy than Bagging, and increasing the AdaBoost rounds monotonically improves performance even at a fixed circuit depth. Moreover, increasing the number of qubits  improves regression performance, suggesting that the model effectively exploits the greater expressive capacity of larger quantum circuits.\\

    Our results suggest that regression performance depends not only on the size and depth of quantum circuits but also on how input features are selected, encoded, and used within an ensemble. These findings provide a promising direction for developing more effective quantum regression methods under practical hardware constraints. Future works could explore more input repetitions  in the encoding part or use trigonometric feature selection to extend the framework toward approximating nonlinear functions in the Fourier domain.  \\
	
	
	\section*{Acknowledgments}
	
	AB acknowledges support from the National Natural Science Foundation of China (grants No.~W2541020, No.~12274059, No.~12574528, and No.~1251101297). MS also acknowledges support from the National Natural Science Foundation of China (grant No.~W2633018). Authors warmly thank Qingyu Li for discussions and valuable suggestions.

	
	\nocite{apsrev42Control} 
	\bibliographystyle{apsrev4-2}
	\bibliography{refs}

\end{document}